\documentclass{aa}  

\usepackage{graphicx}   
\usepackage{txfonts}    
\usepackage{epstopdf} 
\usepackage[colorlinks=true, linkcolor=blue, citecolor=blue, urlcolor=blue, breaklinks=true]{hyperref}
\usepackage{appendix}
\usepackage{xcolor}
\usepackage[normalem]{ulem}

\begin{document} 

  \title{The impact of selection criteria on the properties of \\ green valley galaxies}


   \author{Beatrice Nyiransengiyumva
          \inst{1,2}\fnmsep\thanks{beatny1990@gmail.com}, Mirjana Povi\'c \inst{3,4,1}, Pheneas Nkundabakura \inst{2},\\ Tom Mutabazi \inst{1},
          \and
          Antoine Mahoro \inst{5, 6}}
\authorrunning{B. Nyiransengiyumva}
\titlerunning{Green valley galaxies selection criteria}
   \institute{Department of Physics, Mbarara University of Science and Technology, P.\,O. Box 1410, Mbarara, Uganda
         \and
             MSPE Department, School of Education, College of Education, University of Rwanda, P.\,O. Box 55, Rwamagana, Rwanda
             \and Instituto de Astrof\'isica de Andaluc\'ia (IAA-CSIC), Glorieta de la Astronom\'ia s/n, 18008 Granada, Spain
             \and Astronomy and Astrophysics Department, Entoto Observatory and Research Center (EORC), Space Science and Geospatial\\ Institute (SSGI), P.\,O. Box 33679, Addis Ababa, Ethiopia
    \and South African Astronomical Observatory (SAAO), P.\,O. Box 9, Observatory, Cape Town 7935, South Africa
    \and Southern African Large Telescope (SALT), P.\,O. Box 9, Observatory, Cape Town 7935, South Africa
 }

   \date{Received XXX; accepted XXX}

 
  \abstract
  {\textit{Context:} The bi-modality in the distribution of galaxies usually obtained from colour-colour or colour-stellar mass (absolute magnitude) diagrams has been studied to show the difference between the galaxies in the blue cloud and in the red sequence and to define the green valley region. As a transition region, the green valley galaxies can give clues about morphological transformation of galaxies from late- to early-types, and therefore the selection of green valley is of fundamental importance. \\
  \textit{Aims:} In this work, for the first time, we evaluate the selection effects of the most used green valley selection criteria. The aim is to understand how these criteria affect the identification of green valley galaxies, their properties, and their impact on galaxy evolution studies. \\
  \textit{Methods:} Using the SDSS optical and GALEX ultraviolet data at redshift z\,$<$\,0.1, we selected the eight most commonly used criteria based on colours (without and with Gaussian fittings), specific star formation rate, and star formation rate vs. stellar mass. We then studied the properties of the green valley galaxies (e.g., their stellar mass, star formation rate, specific star formation rate, intrinsic brightness, morphological and spectroscopic types) for each selection criterion.\\
  \textit{Results:} We found that when using different criteria, we select different types of galaxies. UV-optical colour-based criteria tend to select more massive galaxies, with lower star formation rates, with higher fractions of composite and elliptical galaxies, than when using pure optical colours. Our results also show that the colour-based criteria are the most sensitive to galaxy properties, rapidly changing the selection of green valley galaxies. \\
  \textit{Conclusions:} Whenever possible, we suggest avoiding the green valley colour-based selection and using other methods or a combination of several, such as the star formation rate vs. stellar mass or specific star formation rate.}
   
   
   
   {}


   \maketitle
   \nolinenumbers
%

\section{Introduction}
The distribution of galaxies  in colour--stellar mass\,(M$_\ast$), colour--magnitude, colour--star formation rate (SFR), or SFR--M$_\ast$ diagrams show two overdensity regions at both low and high redshift. These overdensity regions are called the red sequence and the blue cloud (e.g., \citealt{Strateva2001, Baldry2004, Salim2007, Brammer2009, Povic2013, Lee2015, Combes2016, Nogueira-Cavalcante2018, Phillipps2019, Sampaio2022, Noirot2022}). In general, the red sequence is mainly populated by quiescent galaxies with early-type morphologies, but it also contains dusty star-forming and edge-on spiral galaxies, while blue cloud mainly hosts star-forming galaxies rich in gas and dust and with late-type morphologies, and a small fraction of early-type galaxies with recent star formation and/or active galactic nuclei (AGN) (e.g., \citealt{Povic2013, Tojeiro2013, Schawinski, Jian2020, Donevski2023, Paspaliaris2023, LeBail2024}). The green valley, located between the red sequence and the blue cloud, is a sparsely populated region with a significantly smaller number of galaxies, at least in the UV-optical-NIR surveys (e.g., \citealt{Salim2014, Bremer2018, Phillipps2019, Jian2020, Noirot2022}). It was suggested that green valley galaxies may represent the transition population between the blue cloud of star-forming galaxies and the red sequence of quenched and passive galaxies \cite[e.g.,][]{Martin2007, Salim2007, Schiminovich2007, Wyder2007, Pan2013, Salim2014, Lee2015, Smethurst2015, Coenda, Angthopo2019, Angthopo2020, Freitas2022, Smith, Estrada-Carpenter2023, Mazzilli2024, Das2025}. In the SFR-M$_{*}$ plane, the green valley can be described as the region located below the main sequence of star formation (e.g., \citealt{Noeske2007, Chang, Ilbert2015, Akiyama2018, Belfiore2018, Jian2020, Sampaio2022, Koprowski2024}).\\
\indent Green valley galaxies account for approximately 10\%\,-\,20\% of galaxies in all environments (e.g., \citealt{Schawinski, Jian2020, Das2021}). In most of the previous studies at both low and intermediate redshifts (z < 1.5), the green valley galaxies represent an intermediate population, with properties (e.g., luminosity, stellar mass, SFRs, stellar ages, stellar populations, etc) between those of red sequence and blue cloud galaxies (e.g., \citealt{Salim2007, Pan2013, Schawinski, Lee2015, Smethurst2015, Trayf, Coenda, Phillipps2019}). They contain intermediate distributions of structural parameters such as S\'ersic index, concentration parameters, asymmetry, smoothness, and bulge-to-total flux ratio (e.g., \citealt{Schiminovich2007, Mendez2011, Mahoro2019}). However, certain inconsistencies have been found in previous works regarding morphologies of green valley galaxies, reporting spirals to be between 70\%\,-\,95\% and 
the elliptical galaxies to account for 5\%\,-\,30\%, depending on the study (e.g., \citealt{Salim2014, Bait2017, Lin2017, Das2021, Aguilar2025}). Inconsistencies have also been found when studying the environments of green valley galaxies, reporting from strong, through mild environmental effects, but also the lack of dependence on the environment. 
For example, some works suggested that stellar mass, SFR, stellar age, the fraction of AGN and the morphology of green valley galaxies are all unaffected by their environments (e.g., \citealt{Schawinski, Starkenburg2019, Jian2020, Das2021}), while others found that the properties of green valley galaxies are environment-dependent  (e.g., \citealt{Coenda}).
Regarding the AGN properties, X-ray and optical studies found a larger fraction of AGN in the green valley (up to z\,$<$\,2), suggesting that AGN feedback may be important for star formation quenching in the green valley (e.g., \citealt{Nandra, Hasinger2008, Silverman2008, Povic2012, Cimatti_2013, Leslie, Lacerda2020, Das2021}). However, when considering FIR emitters in the green valley, enhanced SFRs have been found in active in comparison to non-active galaxies, independently of morphology, suggesting the possible signs of positive AGN feedback (e.g., \citealt{Mahoro2017, Mahoro2019, Mahoro2022, Mahoro2023}).\\
\indent For the morphological transformation of galaxies within the green valley, different timescales have been proposed, from short ($<$\,1\,Gyr), 
intermediate (1\,-\,2\,Gyr), to slow quenching ($>$\,2\,Gyr) \citep[e.g.,][] {Faber2007, Balogh2011, Schawinski, Lee2015, Smethurst2015, Nogueira-Cavalcante2018, Gu2018, Phillipps2019, Angthopo2019, Kacprzak2021, Estrada-Carpenter2023}. In addition, various studies offered a scenario assuming a distinct evolution of cosmic gas supply and gas reservoirs, concluding that early- and late-type galaxies take significantly different evolutionary paths to and through the green valley. It has been suggested that early-type galaxies undergo a rapid quenching of star formation ($<$\,250\,Myr), passing through the green valley as quickly as stellar evolution allows. Late-type galaxies, on the other hand, go through a slow decline in star formation and a gradual departure from the main sequence ($>$\,1\,Gyr) \citep[e.g.,][]{Schawinski, Bremer2018, Kelvin2018, Eales2018, Bryukhareva2019, Wu2020, Quilley_Lapparent2022}. Different processes of star formation quenching have been suggested to play a role in the green valley, from inside-out (e.g., \citealt{Zewdie2020}) to outside-in (e.g., \citealt{Starkenburg2019}).\\
\indent Several selection criteria were previously used to define the green galaxy, including colour-based methods such as U\,-\,V \citep[e.g.,][]{Brammer2009}, U\,-\,B \citep[e.g.,][]{Mendez2011, Mahoro2017, Mahoro2019}, NUV\,-\,r \citep[e.g.,][]{Wyder2007, Salim2007, Salim2014, Lee2015, Coenda, Freitas2022, Noirot2022}, g\,-\,r \citep[e.g.,][]{Trayf, Walker2013, Eales2018}, u\,-\,r \citep[e.g.,][]{Bremer2018, Eales2018, Kelvin2018, Ge2019, Phillipps2019}; SFR-based methods such as: sSFR \citep[e.g.,][]{Schiminovich2007, Salim2009, Salim2014, Phillipps2019, Starkenburg2019}, SFR--M$_{*}$ diagram \citep[e.g.,][]{Noeske2007, Chang, Pandya2017, Jian2020, Sampaio2022}, or D$_\text{n}$4000 index \citep[e.g.,][]{Angthopo2019, Angthopo2020}; or in some cases hybrid methods, which mix the above properties (e.g., colour-colour diagrams, colour vs. sSFR diagram, etc.) \citep[][and references therein]{Salim2014}. Some other criteria based on infrared data, full spectral energy distribution (SED) fitting or star formation quenching indicators have been used in the literature (e.g, \citealt{Mahoro2017, Brownson2020, Noirot2022, Mahoro2023, Villanueva2023}), however, many of them are extensions or refinements of the other criteria mentioned above (e.g., \citealt{Mahoro2017, Mahoro2023}). As can be seen, colour criteria have been the most used in previous studies, mainly by visual selection and, in some cases, by Gaussian fitting of colour distribution. It has been suggested that NUV\,-\,r colour outperforms the colours u\,-\,r and g\,-\,r in the selection of green valley galaxies because the black-body radiation spectra of young stellar populations peak in the NUV, resulting in a higher dynamic range and hence improved separation in the usual colours of red sequence and blue cloud (e.g., \citealp{Gu2018}). However, despite the fact that there are numerous criteria, no study has yet examined their selection effects and tried to understand how and if the inconsistent results obtained in previous studies (and mentioned above) might be related to the green valley selection. This work aims to study for the first time the eight most used green valley selection criteria based on optical SDSS and UV GALEX data (colour without and with Gaussian fitting (g - r, NUV - r), sSFR, and SFR-M$_{\ast}$ criteria), to better understand the properties of the selected galaxies and the impact of tested criteria on the results obtained previously (e.g., \citealt{Salim2009, Walker2013, Salim2014, Trayf, Eales2018, Sampaio2022}, and references therein). These eight criteria are considered the most widely used and to be standard ones, as they appear frequently in the literature (e.g., \citealt{Schawinski, Coenda, Bremer2018}), in particular in all-sky SDSS-GALEX studies at low redshift (e.g., \citealt{Bremer2018, Turner2021}). In selecting these eight criteria, we include different green valley selection methods, as mentioned above (e.g., between 1d and 2d parameter space).\\
\indent From now on, the paper is organised as follows: In Section~\ref{data}, we present the data used. Section~\ref{sample} discusses sample selection and different green valley selection criteria. Section~\ref{analysis} summarises the main analysis and results of this work. In Section~\ref{discussion} we discuss the implications of our results, while the summary and conclusions are presented in Section~\ref{conclusion}. We assume the following cosmological parameters throughout the paper: $\Omega_{m}$ = 0.3, $\Omega_{\Lambda}$ = 0.7, with H$_{0}$ = 70\,km\,s$^{-1}$\,Mpc$^{-1}$. The AB system was used to calculate all magnitudes in this study.

\section{Data}\label{data}
In this work, we used optical and UV data.
In optical, we used the data from Sloan Digital Sky Survey (SDSS) \citep{York2000} Data Release 7 \citep[DR7,][]{Abazajian2009}. In particular, we used the MPA-JHU\footnote{https://wwwmpa.mpa- garching.mpg.de/SDSS/DR7/} catalogue of 927552 sources, with available measurements of emission line fluxes, stellar masses, and SFRs used in Sections~\ref{sample} and ~\ref{analysis}. The extracted stellar masses were measured through a fit to the spectral energy distribution (SED) by using the SDSS broad-band optical photometry \citep{Kauffmann2003}. The H$_{\alpha}$ emission-line luminosity was used to determine the SFRs for the star-forming galaxies as classified in the BPT diagram, and was corrected for dust extinction  \citep{Brinchmann}. For all other galaxies, either AGN or non-emission-line galaxies, the SFRs were inferred from the D$_{4000}$--SFR relation \citep{Brinchmann}. For having a complete sample in the SDSS, we selected all galaxies with redshift z\,$<$\,0.1 \citep{Netzer}. This resulted in a sample of 381798 galaxies, hereafter referred to as the \textbf{"optical sample"}. \\
\indent The UV data were extracted from the GALEX-AIS (GR5) catalogue of 6.5 million sources down to a magnitude of 19.9 and 20.8 in FUV and NUV, respectively \citep{Martin}. We used in particular the NUV data, centred at 2300\,$\AA{}$. \cite{Bianchi2014} found that the SDSS data depth matches well the GALEX-AIS catalogue used in this work. Using a radius of 2\,arcsec, we cross-matched the optical and UV catalogues and obtained a total sample of 179419 galaxies, hereafter referred to as the \textbf{"UV sample"}.

\section{Green valley selection criteria}\label{sample}
In this work, we analysed and compared for the first time the eight commonly used green valley selection criteria. Using both optical and UV samples, we selected colour-based methods, such as g\,-\,r and NUV\,-\,r, with the green valley selected visually and by the Gaussian fitting, and SFR-based methods, such as sSFR and SFR-M$_\ast$. We have selected these eight criteria because they are standard and have been the most used in previous works with SDSS and GALEX data in the low-redshift universe (e.g., \citealt{Schawinski, Coenda,Bremer2018, Turner2021}), however, their reliability and the comparison of the selected galaxies have never been compared before. While additional criteria exist in the literature, their application has been either less common and/or is limited by incomplete data coverage. The number of green valley galaxies selected with each of the tested criteria is given in Table~\ref{table-1}. Detailed information regarding the different selection criteria is provided in Sections \ref{colour-based} -- \ref{gaussian-based} in optically- and UV-selected green valley galaxies.

\begin{table}[ht!]
\caption{Number of green valley galaxies obtained using eight standard selection criteria.}
 \centering
 \begin{tabular}{l c}\hline\hline
  Selection criteria & Number of sources \\
  &(\% of the total sample)\\  \hline
  C1 (g\,-\,r) & 78959 (21\%) \\ 
  C2 (NUV\,-\,r)& 18230 (10\%) \\ 
 C3 (SFR-M$_\ast$, optical)& 54187 (14\%) \\
  C4 (SFR-M$_\ast$, UV) & 17871 (10\%) \\  
 C5 (sSFR, optical) & 84753 (22\%) \\ 
 C6 (sSFR, UV) & 30884 (17\%)\\ 
  C7 (g\,-\,r, Gauss) & 116150 (30\%) \\
C8 (NUV\,-\,r, Gauss) & 55438 (31\%) \\ \hline  

  \end{tabular}
\label{table-1}
\end{table}

\subsection{\textbf{Colour-based methods}} \label{colour-based}
\subsubsection{g\,-\,r criterion (C1)}\label{colour(Opt)}
\begin{figure}
\centering
\includegraphics[width=0.4\textwidth]{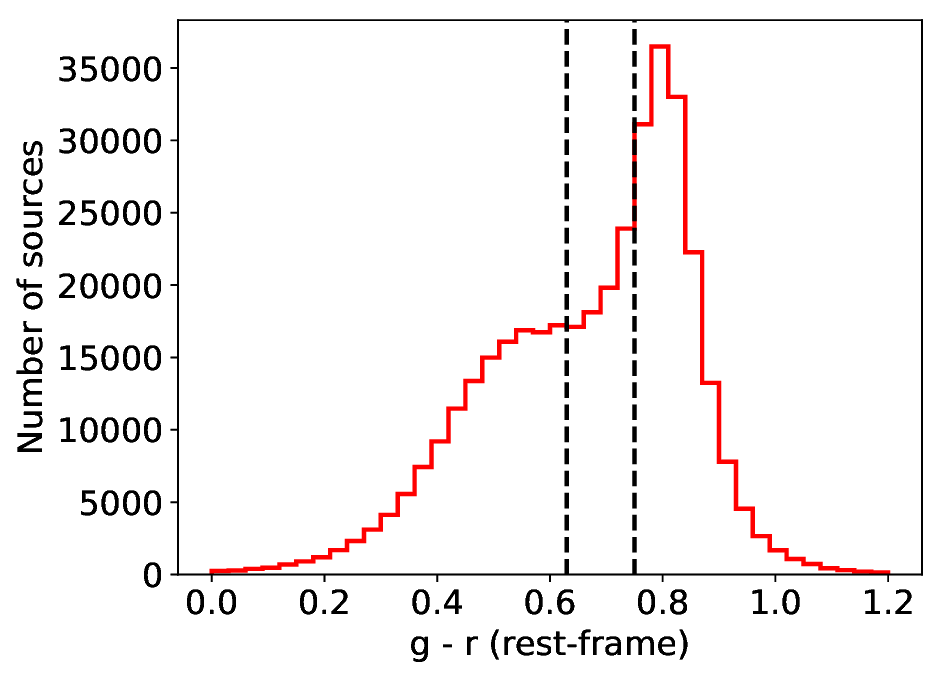}
\caption{Distribution of g\,-\,r rest-frame colour for the optical sample. The two black dashed lines define the green valley region, with the red sequence peak (on the left) and the blue cloud peak (on the right).}
\label{fig-1}
\end{figure}

The first criterion that we selected in our analysis is the g\,-\,r rest-frame colour in optical \citep[hereafter C1; e.g.,][]{Trayf, Walker2013, Eales2018}. We obtained the rest-frame colour through the K-corrected magnitudes using TOPCAT\footnote{https://www.star.bris.ac.uk/~mbt/topcat/}  \citep{Taylor2013, Taylor2017}. 90\% of all sources have $g$ and $r$ magnitude errors below 5\%. The distribution of g\,-\,r rest-frame colour for all galaxies selected in optical is shown in Fig.~\ref{fig-1}. Taking into account the distribution of all galaxies, we defined using purily visual inspection the green valley as the region between the two peaks, corresponding to the blue cloud and the red sequence, with 0.63\,$<$\,g\,-\,r\,$<$\,0.75, as indicated in Fig.~\ref{fig-1} with vertical dashed lines. Using this criterion, 78959 ($\sim$\,21\%) of the 381798 galaxies were selected as green valley. \\
\indent Our definition C1 of green valley is comparable with previous studies. For example, \cite{Trayf} used the galaxies from the EAGLE survey at z\,$<$\,0.1, and by using the same rest-frame colour they defined the green valley as 0.6\,$<$\,g\,-\,r\,$<$\,0.75. \cite{Belfiore2018} using the MaNGA sample defined similarly the green valley as 0.61\,$<$\,g\,-\,r\,$<$\,0.71 in the redshift range of 0.01\,$<$\,z\,$<$\,0.15, and \citet{Walker2013} who used the same criterion defined green valley as 0.55\,$<$\,g\,-\,r\,$<$\,0.70, using a relatively small sample of SDSS galaxies. \\
\indent The literature emphasises that the green valley is generally identified as the region between the two peaks (blue cloud and red sequence) in the colour distribution of galaxies, but the exact boundaries may vary and are often subjective \citep[e.g.,][]{Salim2014, Angthopo2020, Pandey2024}.To clarify, the visual boundaries chosen here are based on the balance between: 1) pure visual inspection and the balance between the general distribution of the two peaks in the colour distribution, and 2) comparison with previous literature, in particular considering the same redshift range for the samples, to ensure that the boundaries are consistent with previous studies. Since the main objective of this work is not to study/propose new selection criteria for green valley galaxies, but to better understand and compare the selection effects and properties of green valley galaxies previously selected using different (and most commonly used) criteria, although our visually determined green valley may be subjective, the boundaries are broadly consistent with previous studies, and consistent therefore with our main objective. However, in order to test/avoid the impact of this subjectivity, we also test more quantitative criteria using Gaussian fitting in sec.~\ref{gaussian-based}.

\subsubsection{\textbf{NUV\,-\,r criterion (C2)}} \label{colour(UV)}

\begin{figure}
\centering
 \includegraphics[width=0.4\textwidth]{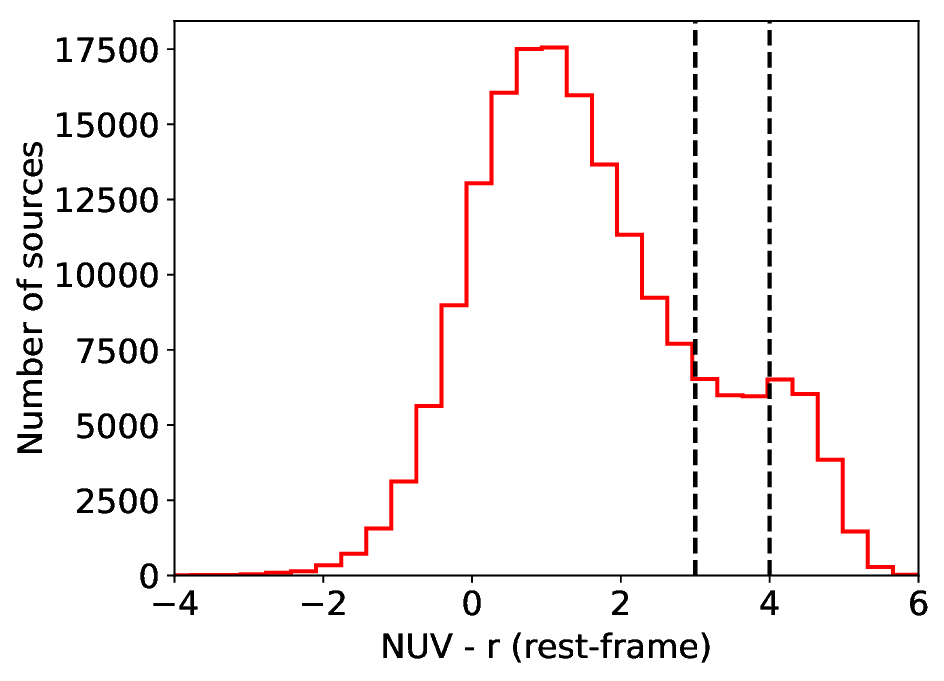}
\caption{Distribution of the NUV\,-\,r colour for the UV sample. The green valley is defined between the two dashed lines.}
\label{fig-4}
\end{figure}

Colours based on the NUV and optical photometric bands are commonly used to select the green valley \citep[e.g.,][]{Lee2015, Coenda, Noirot2022}, and in particular the NUV\,-\,r colour, which is a sensitive tracer of recent star formation activity. Using our UV sample, the NUV\,-\,r bi-modal colour distribution is shown in Fig.~\ref{fig-4}. For the same reason as in Section~\ref{colour(Opt)}, we define the green valley visually, between the peaks of the blue cloud and the red sequence. It is located between the two vertical dashed lines as a region between the two peaks defined as 3\,$<$\,NUV\,-\,r\,$<$\,4. Using this criterion, we obtained 18230 ($\sim$\,10\%) green valley galaxies (out of a total UV sample of 179419 galaxies). \\
\indent Several previous studies have used similar criteria and visual identification to select the green valley, depending on the redshift. \cite{Mendez2011} used a green valley sample selected from the All-Wavelength Extended Growth Strip International Survey (AEGIS) at redshift of z\,$<$\,1.2 using the NUV\,-\,r rest-frame colours and the range of 3.2\,$<$\,NUV\,-\,r\,$<$\,4.1. \cite{Salim2014} used non-dust corrected SDSS and GALEX data at z\,$<$\,0.22, and the green valley was selected within the range of 4\,$<$\,NUV\,-\,r\,$<$\,5 in the UV-optical rest-frame colour, while \cite{Salim2009} selected the green valley using the range of 3.5\,$<$\,NUV\,-\,r\,$<$\,4.5 in the rest frame not-dust corrected colours at the redshift of z\,$<$\,1.4. In addition, \cite{Belfiore2018} used the MaNGA catalogue and defined the green valley galaxies using the UV-optical rest-frame colour and the ranges of 4\,$<$\,NUV\,-\,r\,$<$\,5 at the redshift of 0.01\,$<$\,z\,$<$\,0.15. \cite{Freitas2022} used the SDSS DR12 at the redshift of 0.05\,$\leq$\,z\,$\leq$\,0.095 and defined the green valley in the range of 2.8\,$<$\,NUV\,-\,r\,$<$\,3.5.

\subsection{\textbf{SFR-based methods}}
\subsubsection{SFR-M$_\ast$ criterion in optical (C3)} \label{SFR(Opt)}

\begin{figure}
\centering
\includegraphics[width=0.4\textwidth]{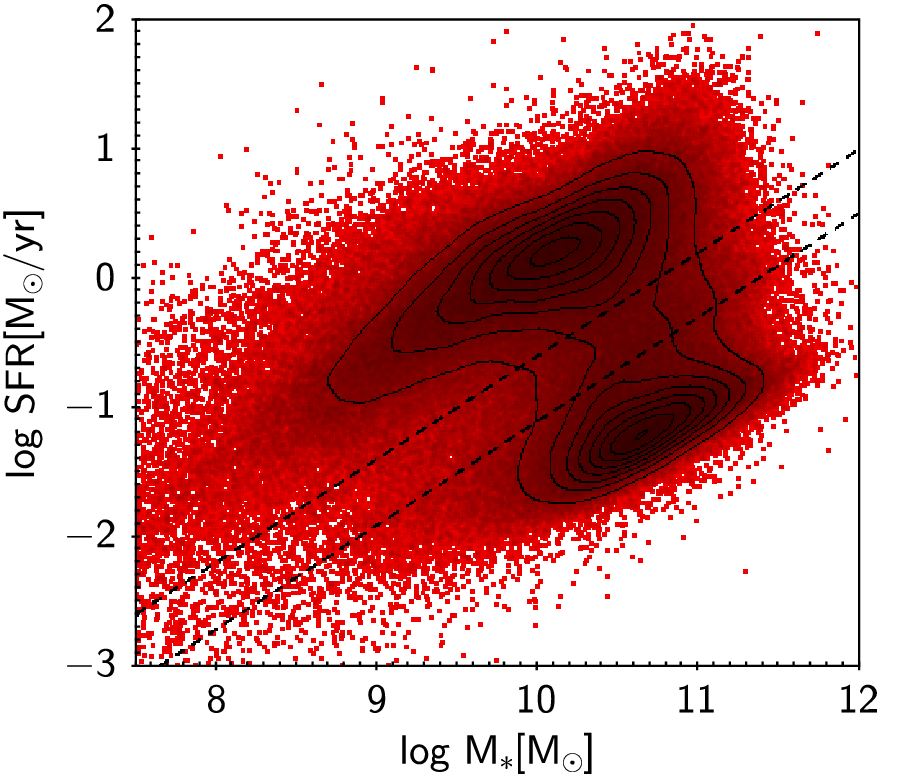}
\caption{SFR-M$_\ast$ diagram for an optical sample of galaxies. The green valley is defined between the two dashed lines. These lines were determined by plotting the density contours in the galaxy distribution.}
\label{fig-3}
\end{figure}

The SFR-M$_\ast$ diagram was also commonly
used to select the green valley galaxies, as a region between the star-forming and passive galaxies without active star formation going on (e.g., \citealp{Chang, Curtis2021, Kalinova2021, Vilella-Rojo2021}). We used the SFR-M$_\ast$ diagram as our third criterion (hereafter C3) to select the green valley galaxies, by plotting the density contours to highlight regions of varying galaxy concentrations where the overdensity contours reveal two prominent peaks corresponding to the star-forming and quiescent populations. The green valley is then defined empirically where the density contours show a minimum in the number density, as shown in Fig.~\ref{fig-3}, and in line with \cite{Schawinski}. With this in mind, we selected the green valley region between the two dashed lines defined as:
\begin{eqnarray}
 \log\,\text{SFR}\,[\text{M}_{\odot}/\text{yr}]&=&0.80 *\log\,(\text{M}_{\ast}[\text{M}_{\odot}]) - 9.155,
 \label{eq1} \\
 \log\,\text{SFR}\,[\text{M}_{\odot}/\text{yr}]&=&0.80 *\log\,(\text{M}_{\ast}[\text{M}_{\odot}]) - 8.599.
 \label{eq2}
 \end{eqnarray}
between the two clear over-density regions of the blue cloud (above the selected green valley region) and the red sequence (below the green valley). Using C3, we selected 54187 ($\sim$\,14\%) green valley galaxies in optical. \\
\indent Our selection and eqs.~\ref{eq1} and \ref{eq2} correspond to several previous studies, including \cite{Belfiore2018}, \cite{Starkenburg2019}, \cite{Jian2020}, and \cite{Sampaio2022}.

\subsubsection{\text{SFR-M$_{*}$ criterion in UV (C4)}} \label{SFR(UV)}

\begin{figure}
\centering
\includegraphics[width=0.4\textwidth]{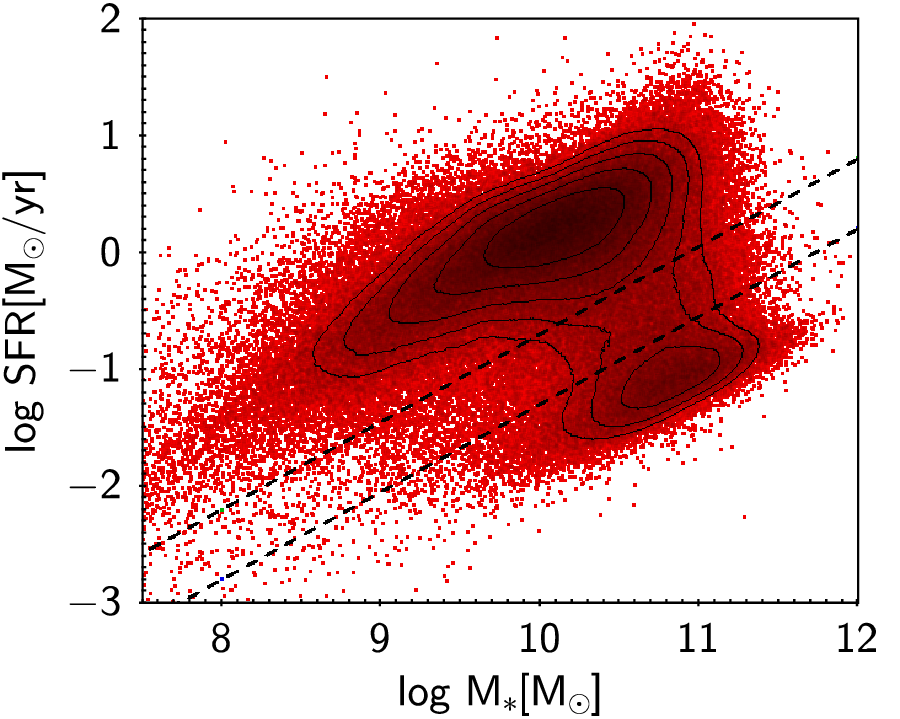}
\caption{SFR versus stellar mass for the UV sample. The green valley is defined between the two dashed lines.}
\label{fig-4b}
\end{figure}

We defined the fourth criterion to our UV sample as SFR-M$_\ast$ to determine the green valley region, as shown in Fig.~\ref{fig-4b}. We followed the same procedure as in Section~\ref{SFR(Opt)} and defined the green valley using the density contours as an area between the two overpopulated regions of the blue cloud and the red sequence with a minimum in number density, marked between the dashed lines using eqs.~\ref{eq3_1} and \ref{eq4_1} as:
%
 \begin{eqnarray}
 \log\,\text{SFR}\,[\text{M}_{\odot}/\text{yr}]&=&0.75 *\log\,(\text{M}_{\ast}[\text{M}_{\odot}]) - 8.88,
 \label{eq3_1} \\
 \log\,\text{SFR}\,[\text{M}_{\odot}/\text{yr}]&=&0.75 *\log\,(\text{M}_{\ast}[\text{M}_{\odot}]) - 8.2.
 \label{eq4_1}
 \end{eqnarray}
 %
 %
We obtained 17871 (10\%) green valley galaxies in total using this criterion. Our selection is in line with previous studies (e.g., \citealp{Noeske2007, Lee2015}).

\subsection{\textbf{sSFR-based methods}}
\subsubsection{\textbf{SFR criterion in optical (C5)}} \label{sSFR(Opt)}

\begin{figure}
\centering
\includegraphics[width=0.4\textwidth]{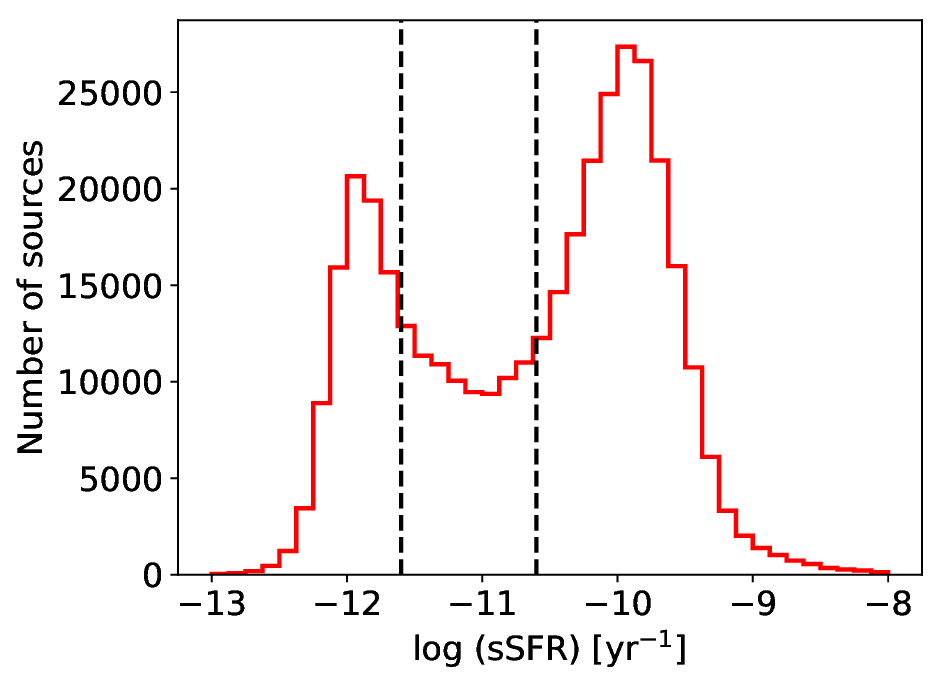}
\caption{sSFR distribution for the optical sample. The green valley lies between the two vertical dashed lines.}
\label{fig-2}
\end{figure}

The sSFR was used commonly in previous works to select the green valley galaxies \citep[e.g.,][]{Schiminovich2007, Salim2009, Salim2014, Phillipps2019, Starkenburg2019}. The distribution of sSFR for the optical sample is shown in Fig.~\ref{fig-2}. In line with the justification given in Sec.~\ref{colour-based}, the green valley sample was selected visually between the two peaks and is represented by the black dashed lines, covering the range of -11.6\,$<$\,sSFR\,$<$\,-10.6. Using this criterion (hereafter C5), we selected a total of 84753 ($\sim$ 22\%) green valley galaxies. \\
\indent Our selection is consistent with previous studies based
on sSFR distribution and visual selection. In \cite{Salim2014}, using also the SDSS data at z\,$<$\,0.22, the green valley was selected as -11.8\,$<$\,log(sSFR)\,$<$\,-10.8, while the range of -11.0\,$<$\,log(sSFR)\,$<$\,-10.0 was used in \cite{Salim2009}. \cite{Phillipps2019} used the data from the GAMA survey at a redshift of 0.1\,$<$\,z\,$<$\,0.2 and defined the green valley using sSFR in the range of -11.5\,$<$\,log(sSFR)\,$<$\,-10.6. On the other hand, using Illustris\footnote{https://www.illustris-project.org/} and EAGLE\footnote{https://eagle.strw.leidenuniv.nl/} simulations at z\,$=$\,0, \cite{Starkenburg2019} defined the green valley galaxies using the sSFR in the range of -13.0\,$<$\,sSFR\,$<$\,-10.5.

\subsubsection{\textbf{sSFR criterion in UV (C6)}} \label{sSFR(UV)}

\begin{figure}
\centering
\includegraphics[width=0.4\textwidth]{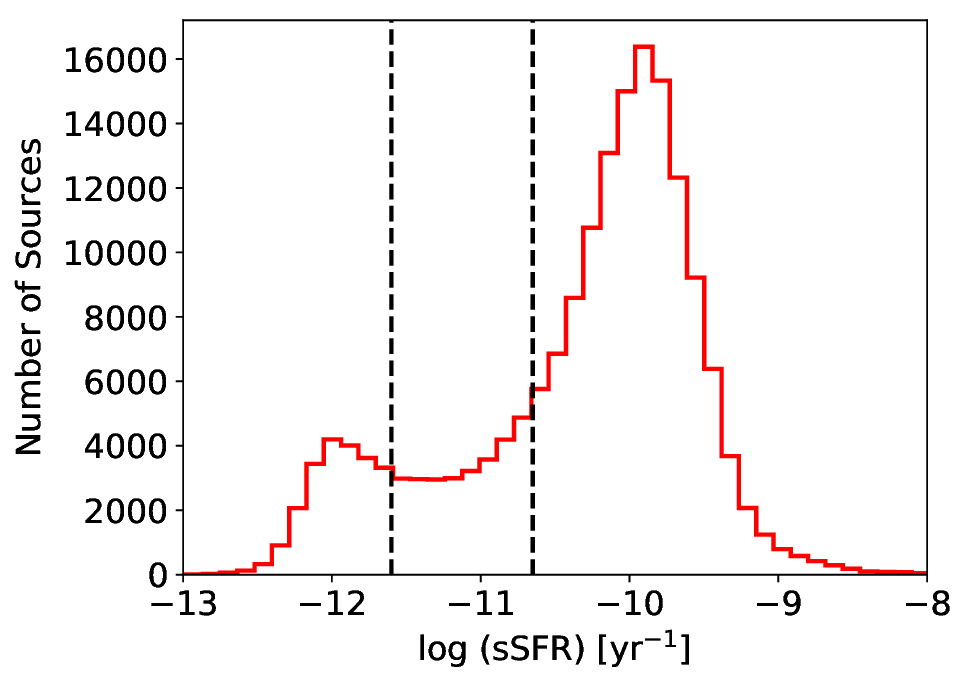}
\caption{Distribution of the sSFR in the UV sample. The green valley is defined between the two dashed lines.}
\label{fig5b}
\end{figure}

The selection of green valley galaxies using the sSFR in UV is similar to the one in optical described in Section~\ref{sSFR(Opt)} and is shown in Fig.~\ref{fig5b}. Using the same criterion visually identified as in optical, we selected 30884 green valley galaxies ($\sim$\,17\%). In \citet{Salim2009} and \citet{Salim2014}, the authors used the same selection criteria, suggesting that the green valley selection using the sSFR does not change significantly independently of the colour used. 

\subsection{\textbf{Gaussian-fitting based methods}}\label{gaussian-based}
\subsubsection{Gaussian fitting on g\,-\,r colour distribution criterion (C7)} \label{colour(grgaussian)}

\begin{figure}
\centering
\includegraphics[width=0.4\textwidth]{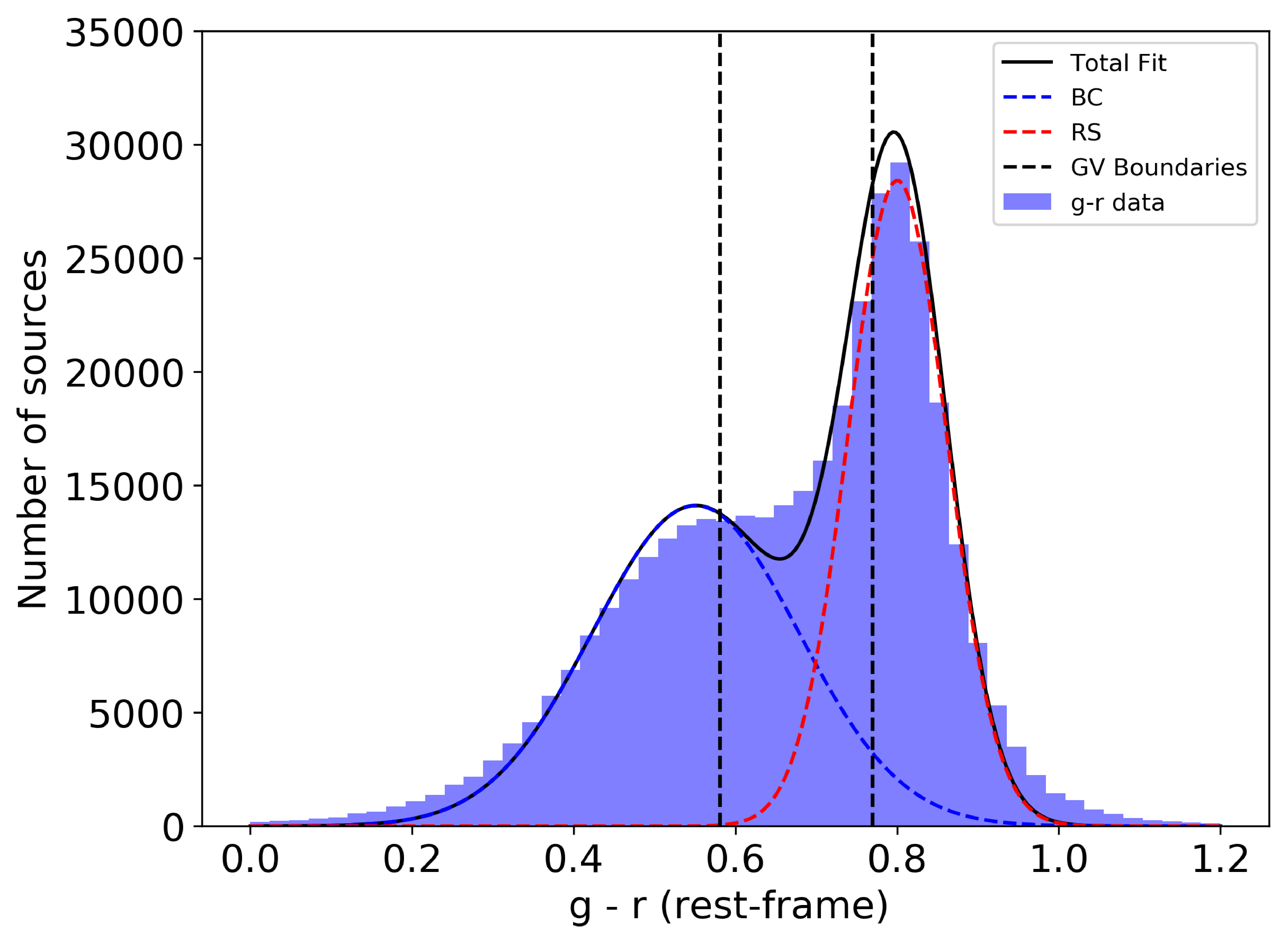}
\caption{The g\,-\,r colour and simultaneous double Gaussian fit criterion. The black solid line represents the best fit. The blue- and red dashed Gaussians show individual components which correspond to the blue cloud and the red sequence, respectively. The green valley region is shown between the two dashed black vertical lines.}
\label{fig1g}
\end{figure}

In order to avoid visual selection, several previous works used the two Gaussian functions to simultaneously fit the distributions of the blue cloud and the red sequence when different colours are used and to define the galaxies in the green valley between the two Gaussian peaks \cite[e.g.,][]{Baldry2004, Wyder2007, Mendez2011, Elisabeth2012, Parente2025A}. In this work, in order to compare  visually and not-visually selected green valley galaxies in the colour-based methods, we performed an additional analysis for the g\,-\,r criterion by simultaneously fitting the bi-modal distribution with two Gaussian components via a simple least-squares fitting. After obtaining the best-fit parameters, we de-blended the Gaussian components to obtain the two individual Gaussians corresponding to the blue cloud and red sequence. This approach is similar to the criterion used by \citet{Parente2025A} when fitting u\,-\,r optical colours, where they defined the green valley to be in the region within $\mu_B + \sigma_B$ to $\mu_R - \sigma_R$ (subscripts $B$ and $R$ refer to the blue cloud and red sequence, respectively), where $\mu$ and $\sigma$ represent the mean and standard deviation, respectively, of each Gaussian fit. However, among other conditions, their definition of the green valley requires that $\mu_R - \sigma_R \ge x_{BR}$ and $\mu_B + \sigma_B \le x_{BR}$, where $x_{BR}$ is the intersection of the two Gaussian components. This method uses Gaussian fits to the observed color distribution, which is a statistical approach to modeling the underlying populations, the means and standard deviations are derived from the data, and the intersection is mathematically determined. Since using a single mean and standard deviation was not selecting properly the green valley in our sample, we modified and redefined the green valley as the region falling within $\frac{1}{2}\left[\left(\mu_R + \mu_B\right) + \left(\sigma_B + \sigma_R\right)\right]$ and $\frac{1}{2}\left[\left(\mu_R + \mu_B\right) - \left(\sigma_B + \sigma_R\right)\right]$, using the average values of $\mu$ and $\sigma$ of the two Gaussian fits instead of using a single mean and standard deviation. Fig.~\ref{fig1g} shows the double-component Gaussian fit (black solid line), the individual components for the blue cloud (blue dashed line) and the red sequence (red dashed line). The vertical black dashed lines define the green valley region in g\,-\,r colour between the values of 0.6-0.77, similar to what we have in Section~\ref{colour(Opt)}. Following this criterion, we selected 116150 ($\sim$\,30\%) green valley galaxies. 

\subsubsection{\textbf{Gaussian fitting on NUV\,-\,r colour distribution criterion (C8)}} \label{colour(UV-gauss)}

\begin{figure}
\centering
\includegraphics[width=0.4\textwidth]{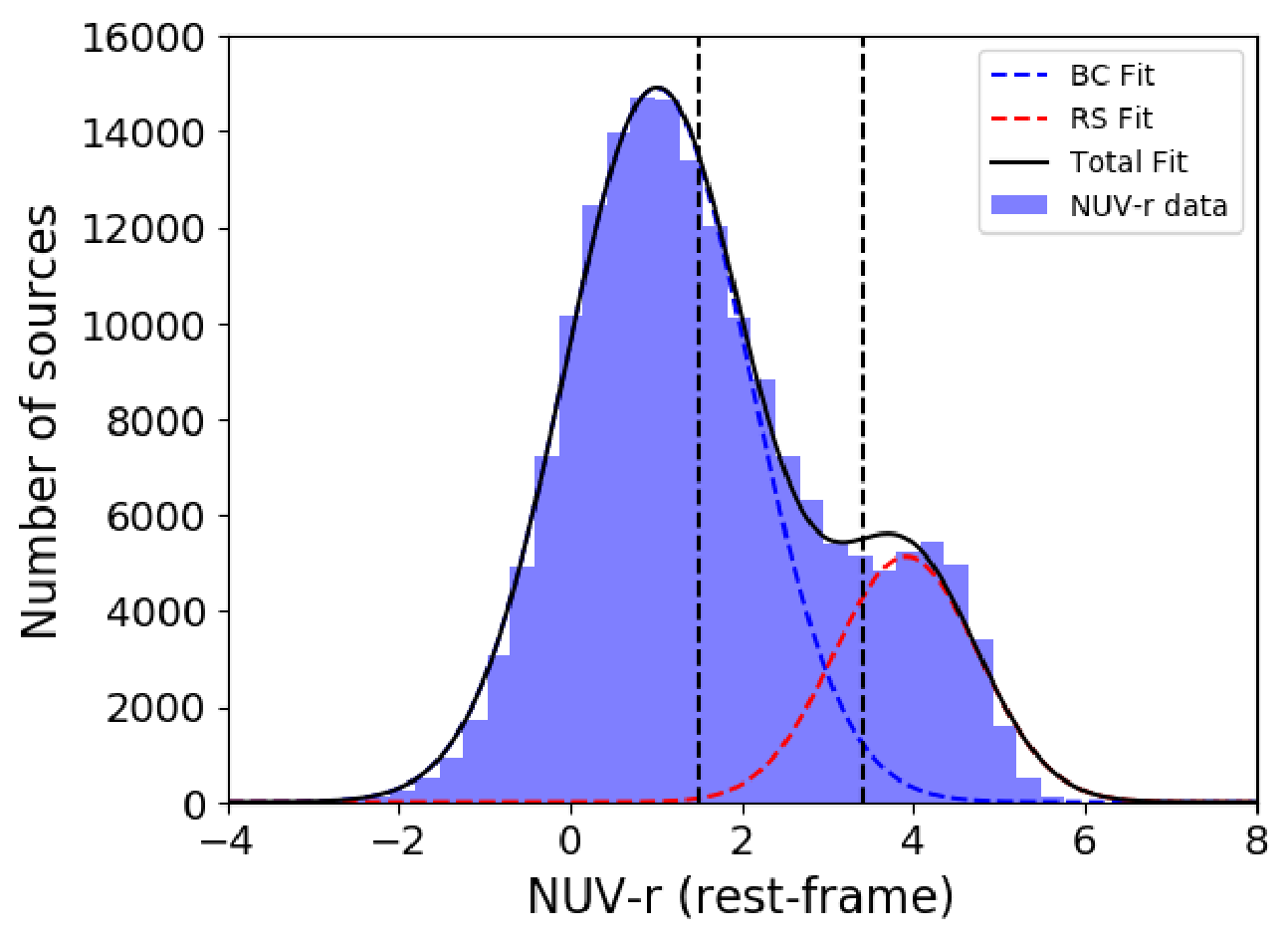}
\caption{The distribution of the NUV\,-\,r colour and Gaussian fittings of the blue cloud (blue dashed line) and the red sequence (red dashed line). The green valley is defined between the two black dashed vertical lines.}
\label{fig2g}
\end{figure}

Similar as in Section~\ref{colour(grgaussian)}, to avoid the subjectivity associated with visually identified green valley galaxies, a Gaussian fit to the NUV\,-\,r colour distribution was performed. We defined the green valley region by fitting the bi-modal distribution simultaneously with two Gaussian components as described in Section~\ref{colour(grgaussian)}. Fig.~\ref{fig2g} shows the double-component fit (black solid line) and the individual components of the blue cloud (blue dashed line) and the red sequence (red dashed line). The green valley region is defined between the two vertical black dashed lines covering the range of values in the NUV\,-\,r colour of 1.5-3.4. This resulted in a sample of 55438\,($\sim$\,31\%) green valley galaxies. 

\section{Data analysis and results}\label{analysis}

In this section, we present some of the key properties of green valley galaxies when selected using the eight different criteria described in Sections~\ref{colour-based}--\ref{gaussian-based}, including the stellar mass, SFRs, sSFRs, intrinsic brightness, morphology and spectroscopic type of galaxies. We also discuss the possible impact of extinction on our results. 


\subsection{Stellar mass}

\begin{figure}[ht!]
\includegraphics[width=0.45\textwidth]{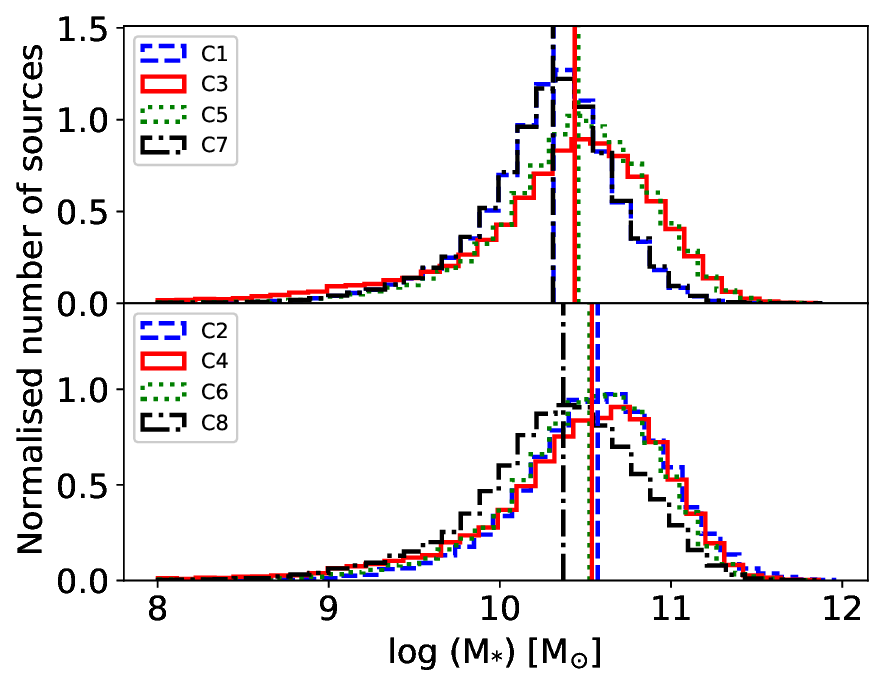}
\caption{Comparison of stellar mass for optically- (top panel) and UV-selected (bottom panel) green valley galaxies using colour (C1 (g\,-\,r) and C2 (NUV\,-\,r), blue dashed histograms), SFR-M$_\ast$ (C3 and C4, red solid histograms, sSFR (C5 and C6, green dotted histograms), and colour with Gaussian fit (C7 (g\,-\,r)  and C8 (NUV\,-\,r), black dot-dashed histograms)) criteria. The vertical dashed lines in both panels represent the median values of each used criterion.}
\label{fig-6}
\end{figure}

The stellar mass distribution of all optically- (top panel) and UV-selected (bottom panel) green valley galaxies is shown in Fig.~\ref{fig-6}. The median values (vertical lines in both figures) of log\,(M$_\ast$) are 10.31, 10.43, 10.45, 10.31,\,[M$_{\odot}$] for optically-selected green valley galaxies using criteria C1, C3 , C5 and C7 respectively, and 10.57, 10.53, 10.52, 10.36\,[M$_{\odot}$] in the UV using criteria C2,  C4, C6 and C8,respectively. The Q1--Q3\footnote{The first quartile (Q1) and third quartile (Q3) give 25\% and 75\% of data points, respectively.} values for all criteria are given in Table~\ref{table22}. As can be seen in Fig.~\ref{fig-6}, the stellar mass distributions in optical are similar for C1 and C7, being lower than in C3 and C5. Similar is the case in UV, with now slightly lower masses obtained for C8 than for C2, C4 and C6. In general, the differences in both optical and UV between the four selection criteria are on average of
order of up to log\,(M$_\ast$)\,=\,$\pm$\,0.14\,[M$_{\odot}$] in optical, and of up to log\,(M$_\ast$)\,=\,$\pm$\,0.2\,[M$_{\odot}$] in UV (between C8 and other three UV criteria). On the other hand, when comparing colour (without and with Gaussian fit), sSFR, and SFR-M$_\ast$ criteria between optical and UV in Fig.~\ref{fig-6} and statistics given in Table~\ref{table22}, it can be seen that the UV green valley criteria tend to select slightly more massive galaxies. The difference is most significant in the colour criterion case (without the Gaussian fitting) compared to other criteria. For this criterion, in optical, 50\% of the sources occupy the log M$_\ast$ range of 10.07\,-\,10.51\,[M$_{\odot}$] with a mean of 10.31\,[M$_{\odot}$], compared to 50\% of the galaxies in the green valley in UV which lie between 10.28\,-\,10.83\,[M$_{\odot}$] and have a mean of 10.57\,[M$_{\odot}$]. These differences become less significant in the case of other three criteria, as can be seen in Fig.~\ref{fig-6} and Table~\ref{table22}. An analysis of the two-sample Kolmogorov-Smirnov (KS) test
for the stellar masses of different optically and UV-selected samples indicates that, in all cases, the two samples do not come from the same parent distribution (p-value $<$\,0.05). The difference is more pronounced for the colour criterion (without Gaussian fitting)
compared to the other criteria, where the D-parameter\footnote{In the KS test, D is the maximum vertical distance between the two comparable distributions. It can take values between 0 (no difference) and 1 (100\% difference).} is 29\%, compared to 11\%, 6.5\%, and 9.2\% for criteria C7 vs. C8, C5 vs. C6, and C3 vs.C4, respectively.

\begin{table}[ht!]
\caption{The log\,(M$_\ast$) statistics in units of [M$_{\odot}$] in the eight green valley criteria.}
\small{
 \centering
 \begin{tabular}{l c c c }\hline\hline
Selection criteria & Q1 &Median& Q3 \\  \hline
 C1 (g\,-\,r)  &  10.07 &10.31 &10.51 \\ 
 C2 (NUV\,-\,r) &  10.28 & 10.57 &10.83 \\
 C3 (SFR-M$_\ast$, optical) &10.08  & 10.43& 10.72 \\
 C4 (SFR-M$_\ast$, UV)&  10.19 &10.53 &10.82  \\
 
 C5(sSFR, optical)&  10.17 & 10.45 &10.72 \\
 C6 (sSFR, UV)& 10.22& 10.52&10.78\\ 
 C7 (g\,-\,r, Gauss) &10.06  & 10.31&10.52 \\

 C8 (NUV\,-\,r, Gauss)&  10.06 & 10.36 &10.65  \\
  \hline\hline
 
  \end{tabular}}
\label{table22}
\end{table}

\subsection{Star formation rates}
\label{sec_SFR}

\begin{figure}[ht!]
\centering
\includegraphics[width=0.45\textwidth]{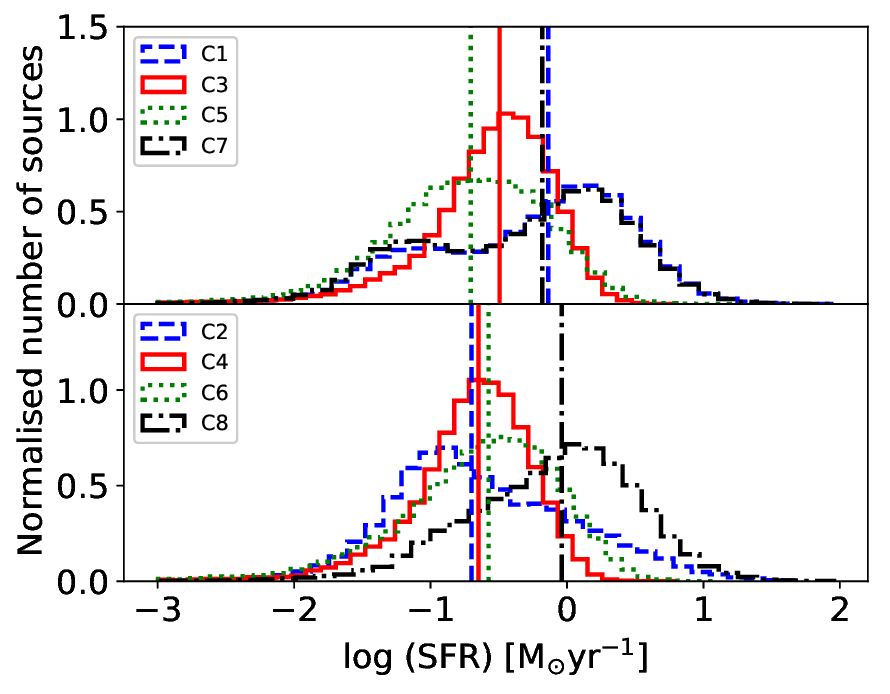}
\caption{Same as Fig.~\ref{fig-6}, but for SFR.}
\label{fig-7}
\end{figure}

Figure~\ref{fig-7} shows the distribution of the SFRs of all green valley samples and their corresponding median values. As in the previous section, when comparing the four criteria in the optical, C1 and C7 show similar distributions, with larger SFRs than C3 and C5 (see Table~\ref{table20c}). On the other hand, for the UV sample, C8 shows a different distribution, with higher SFRs than C2, C4, and C6. As in the case of the stellar mass, the colour criteria without Gaussian fit show the largest difference between the optically-selected and the UV-selected green valley galaxies. When using the C1 criterion, the galaxies show a median log\,(SFR) of $-0.13$\,[M$_{\odot}$yr$^{-1}$] where 50\% of the sources lie between $-0.83$\,-\,$0.26$\,[M$_{\odot}$yr$^{-1}$], compared to the C2 criterion where green valley galaxies show lower values of log\,(SFR) with 50\% of galaxies being in the range of $-1.07$\,-\,($-0.15$)\,[M$_{\odot}$yr$^{-1}$] with a median of $-0.69$\,[M$_{\odot}$yr$^{-1}$]. Differences are less significant for the other three criteria, as can be seen in Table~\ref{table20c}. The two distributions are again the most similar when sSFR criteria are used. As is the case with stellar mass, the two-sample KS test showed that the two samples, optical and UV, do not come from the same parent distribution (p-value $<$ 0.05) in case of all criteria, with the more pronounced difference in colour criteria (without Gaussian fitting) of D\,=\,27\%. 

\begin{table}[ht!]
\caption{The log\,(SFR) statistics in units of [M$_{\odot}$yr$^{-1}$] in the eight green valley criteria.}
\small{
 \centering
 \begin{tabular}{l c c c}\hline\hline
   Selection criteria & Q1 &Median& Q3 \\  \hline
 C1 (g\,-\,r)   &  -0.83 &-0.13 &0.26 \\ 
 C2 (NUV\,-\,r)&  -1.07 & -0.69 &-0.15 \\
 C3 (SFR-M$_\ast$, optical)&-0.78 & -0.49&-0.24 \\
   C4 (SFR-M$_\ast$, UV)&  -0.92 & -0.64 &-0.40 \\
   C5 (sSFR, optical)& -1.09 & -0.70 &-0.32\\
    C6 (sSFR, UV) & -0.97& -0.57&-0.23\\ 
  C7 (g\,-\,r, Gauss)&-0.91 & -0.18&0.24 \\

 C8 (NUV\,-\,r, Gauss)&  -0.49 & -0.03 &0.32 \\

  \hline\hline

  \end{tabular}}
\label{table20c}
\end{table}

\subsection{Specific SFRs}\label{sec_sSFR}

\begin{figure}[ht!]
\centering
\includegraphics[width=0.45\textwidth]{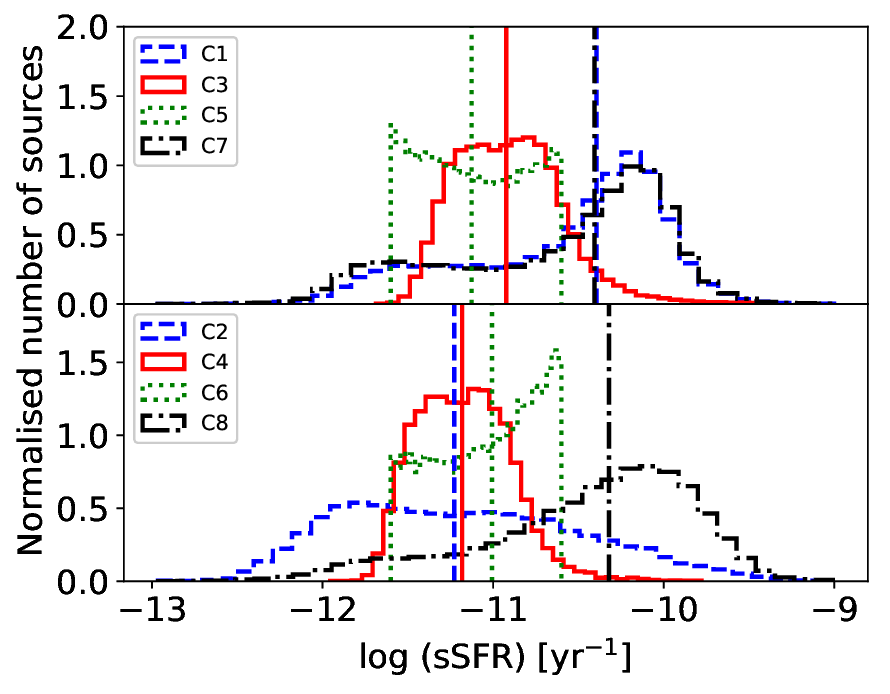}
\caption{Same as Fig.~\ref{fig-6}, but for sSFR.}
\label{fig-8}
\end{figure}

We tested the log\,(sSFR) of selected green valley samples by analysing the distribution of every sample and measuring their statistics. Fig.~\ref{fig-8} shows the distribution of sSFR in optical (top panel) and UV (bottom panel) samples and their corresponding medians. The medians of log\,(sSFR) in our samples are -10.39, -10.92, -11.12 and -10.40 [yr$^{-1}$] for the green valley selected in optical using C1, C3, C5 and  C7 criteria, respectively, and -11.22, -11.18, -11.00 and -10.32 [yr$^{-1}$] for the green valley selected in UV using C2, C4, C6 and C8 criteria, respectively. When comparing the four criteria in the selected optical and UV samples, we find similar trends as in the case of stellar mass and SFRs, with C1 and C7 having similar distributions compared to C3 and C5 in the optical, and with the largest differences between C8 and the other three criteria in the UV. When comparing the same criteria between the optical and UV samples, the difference is again the most significant for the colour criteria without Gaussian fitting, with the median of -10.39 (and Q1-Q3 range of $-10.96$\,-\,$-10.14$) for C1 and -11.22 (Q1-Q3 range of $-11.74$\,-\,$-10.67$) for C2. These differences are less significant in the case of the other three criteria, as can be seen in Table~\ref{table30c}. The results from the two-sample KS test revealed that in all cases, when comparing the same criteria in the optical and UV, the two samples do not come from the same parent distribution (p-value $<$ 0.05), with the colour criterion (no Gaussian fit) again showing the largest difference, with D\,=\,40\%, compared to other criteria where D is lower (12\%, 10\%, 32\% for C7 vs. C8, C5 vs. C6, and C3 vs. C4, respectively).

\begin{table}[ht!]
\caption{log\,(sSFR) statistics in units of [yr$^{-1}$] in the eight green valley criteria.}
\small{
 \centering
 \begin{tabular}{l c c c}\hline\hline
  Selection criteria & Q1 &Median& Q3 \\  \hline

 C1 (g\,-\,r)  &  $-10.96$&$-10.39$ &$-10.14$ \\
  C2 (NUV\,-\,r)&  -11.74 & -11.22 &-10.67\\
  C3 (SFR-M$_\ast$, optical)& -11.14  & -10.92&-10.71 \\
  C4 (SFR-M$_\ast$, UV)& -11.38 & -11.18 &-10.98 \\
  C5 (sSFR, optical)& -11.37& -11.12 &-10.84\\ 
  C6 (sSFR, UV) & -11.29& -11.00&-10.77\\
 C7 (g\,-\,r, Gauss)& -11.10  & -10.40&-10.12 \\
 C8 (NUV\,-\,r, Gauss)& -10.78  & -10.32&-9.9 \\
  \hline\hline
 
  \end{tabular}}
\label{table30c}
\end{table}

\subsection{Absolute magnitude, M$_r$}
\label{sec_mag_abs}

\begin{figure}[ht!]
\centering
\includegraphics[width=0.45\textwidth]{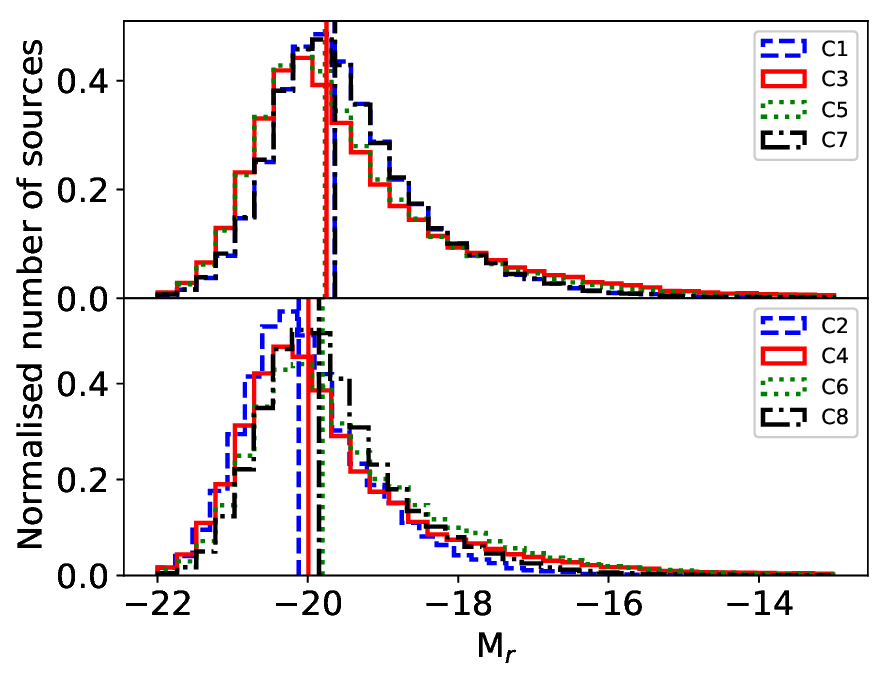}
\caption{Same as Fig.~\ref{fig-6}, but for the absolute magnitude in the r-band.}
\label{figR}
\end{figure}

We analysed the absolute magnitude in r-band (M$_r$) to compare the intrinsic brightness of the optical and UV samples. The median values of M$_r$ are -19.64, -19.75, -19.77 and -19.63 for the optical sample using C1, C3, C5 and C7 criteria, respectively, and -20.12, -19.99, -19.80 and -19.85 for the C2, C4, C6 and C8 criteria, respectively, in the UV. Fig.~\ref{figR} represents the distribution of absolute magnitude in the r-band (M$_{r}$) obtained in all green valley selected optical (top) and UV (bottom) samples. The main statistics of all distributions are provided in Table~\ref{table33c}. We observe that green valley galaxies selected in the UV are intrinsically slightly brighter than galaxies selected using optical criteria. Similar as in the above cases, the KS test showed that the two samples are not from the same parent distribution (all p-values are $<$\,0.05). As is the case with the previously discussed parameters, the difference is more pronounced in colour criteria without any Gaussian fitting (D\,=\,23\%) compared to other criteria where D is 10\%, 2\%, and 9\% for colour with Gaussian fit, sSFR, and SFR vs. M$_\ast$, respectively.

\begin{table}[ht!]
\caption{M$_{r}$ statistics in the eight green valley criteria.}
\small{
 \centering
 \begin{tabular}{l c c c}\hline\hline
  Selection criteria & Q1 &Median& Q3 \\  \hline
C1 (g\,-\,r)& $-20.16$&$-19.64$ &$-18.94$ \\
C2 (NUV\,-\,r)&  -20.58 & -20.12 &-19.52\\
C3 (SFR-M$_\ast$, optical)& -20.34  & -19.75& -18.81 \\
 C4 (SFR-M$_\ast$, UV)& -20.52 & -19.99 &-19.16 \\
 C5 (sSFR, optical)& -20.33 & -19.77 &-18.92 \\
 C6 (sSFR, UV)& -20.38& -19.80&-18.88\\
C7 (g\,-\,r, Gauss)& -20.17  & -19.63& -18.92 \\ 
 C8 (NUV\,-\,r, Gauss)& -20.34  & -19.85& -19.21 \\

  \hline\hline
 
  \end{tabular}}
\label{table33c}
\end{table}


\subsection{Morphological types}

\begin{figure}[ht!]
 \includegraphics[width=0.4\textwidth]{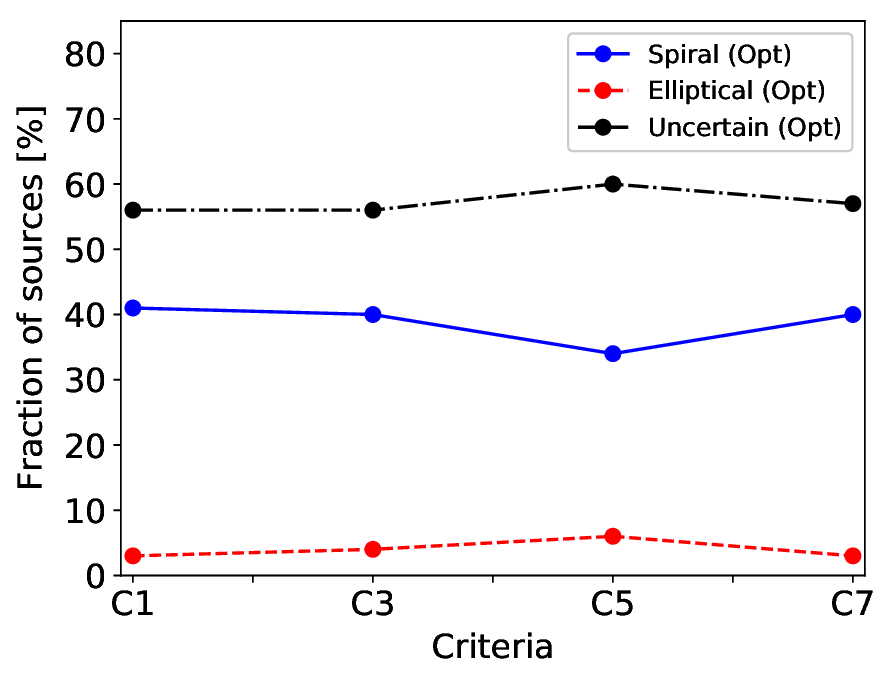}\\
\includegraphics[width=0.4\textwidth]{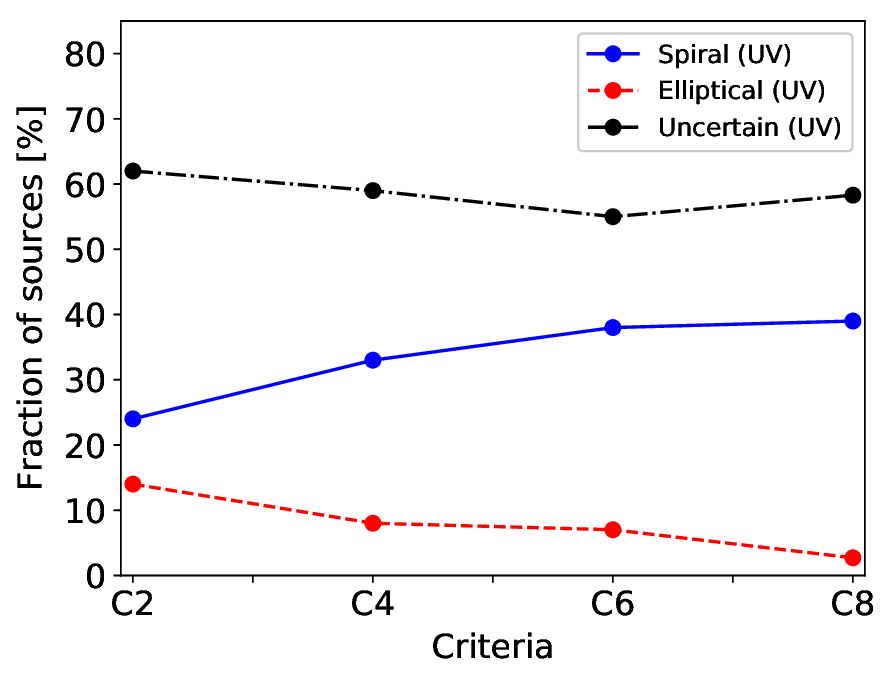}
\caption{Comparison of morphological classification of green valley samples for every criterion ranging from C1, C3, C5 and C7 for optically-selected sample (top panel) and C2, C4, C6 and C8 for UV-selected sample (bottom panel). The morphological types are from the Galaxy Zoo survey. The blue solid line, red dotted line, and the black dash-dotted line represent spiral, elliptical, and uncertain (unclassified) galaxies, respectively.}
\label{fig-10}
\end{figure}

The morphology is a powerful indicator of a galaxy's dynamical and merger history, it plays a crucial role in understanding the evolutionary pathways of green valley galaxies and strongly correlates with many physical parameters, including mass, star formation, and star formation history \citep[e.g.,][]{Schawinski, Povic2012, Smith, Estrada-Carpenter2023}. In this section, we analysed the morphological classification of all green valley galaxies selected using different criteria.\\
\indent We used the visual morphological classification from the Galaxy Zoo citizen science project \cite{Lintott2008, Lintott} where galaxies were classified as: spiral, elliptical and uncertain based on their structure. The "uncertain" classification includes galaxy mergers and interactions, edge-on galaxies, peculiar galaxies and unknown morphologies. Galaxies categorised as "elliptical" or "spiral" require that 80\% of Galaxy Zoo users to have classified it in that category, after the debiasing technique has been carried out. All remaining galaxies are then categorised as uncertain (the \textsc{CLEAN} technique; \citealt{Land2008}). After cross-matching the Galaxy Zoo catalogue with our catalogues of green valley optical and UV samples, we obtained the morphological classifications for each criterion, shown in Fig.~\ref{fig-10}. Table~\ref{table-2} shows the statistics in \% of all morphological types, where the percentages in column 2 are obtained by comparing the number of galaxies after cross-matching with Galaxy Zoo and the number of galaxies in the original green valley samples.\\
\indent As can be seen in both Fig.~\ref{fig-10} and Table~\ref{table-2}, we detect in all criteria more spiral than elliptical galaxies in the green valley. This is not surprising since most of the previous morphological studies (at both low and intermediate redshifts) confirmed that the population of late-type galaxies is higher than that of early-types (see e.g., \citealp{Nair, Povic2013}). On the other hand, the system of classification, together with the quality of SDSS images, is responsible for an excessive number of uncertain sources. When comparing the fractions of elliptical and spiral galaxies in the optical and UV samples, the only significant difference is between the two colour criteria without Gaussian fitting (C1 vs. C2), where we selected more spiral galaxies (41\% vs. 24\%, respectively) than elliptical (3\% vs. 14\%, respectively) in optical and UV, as can be seen in Table~\ref{table-2}).

\begin{table}[ht!]
\caption{Morphological classification using Galaxy Zoo \citep{Lintott2008, Lintott} (Sp for spiral, Ell for elliptical and Unc for uncertain).}
\scriptsize{
 \centering
 \small{
 \begin{tabular}{l c c c c} \hline\hline
 Criteria  &  Parent  & Sp & Ell   & Unc  \\ 
 \hline
  C1 (g\,-\,r)&  70948 (90\%)& 41\% & 3\% & 56\%\\ 
  C2 (NUV\,-\,r) & 17719 (97\%)&24\% & 14\% &62\% \\
  C3 (SFR-M$_\ast$, optical) & 49034 (90\%)& 40\% &4\% &56\%\\
  C4 (SFR-M$_\ast$, UV) &  17299 (96.8\%)&33\% &8\% & 59\%    \\
   C5 (sSFR, optical) & 72148 (85\%)& 34\% & 6\% &60\% \\
    C6 (sSFR, UV) & 27353 (88.5\%)& 38\% & 7\% & 55\%\\
   C7 (g\,-\,r, Gauss) & 104275 (89.7\%)& 40\% &3\% &57\%\\
 C8 (NUV\,-\,r, Gauss) & 51796 (93.5\%)& 33\% &7\%&60\%\\
    \hline\hline
 \end{tabular}}}
 \label{table-2}
 \end{table}
 
 \subsection{Spectroscopic types}
 
\begin{figure}[ht!]
\centering
 \includegraphics[width=0.4\textwidth]{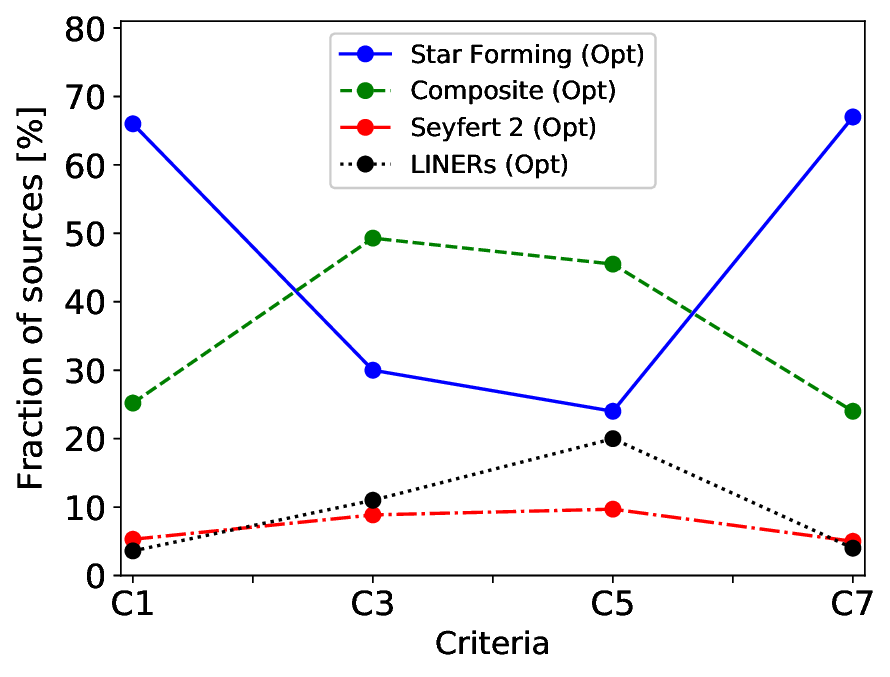}\\
 \includegraphics[width=0.4\textwidth]{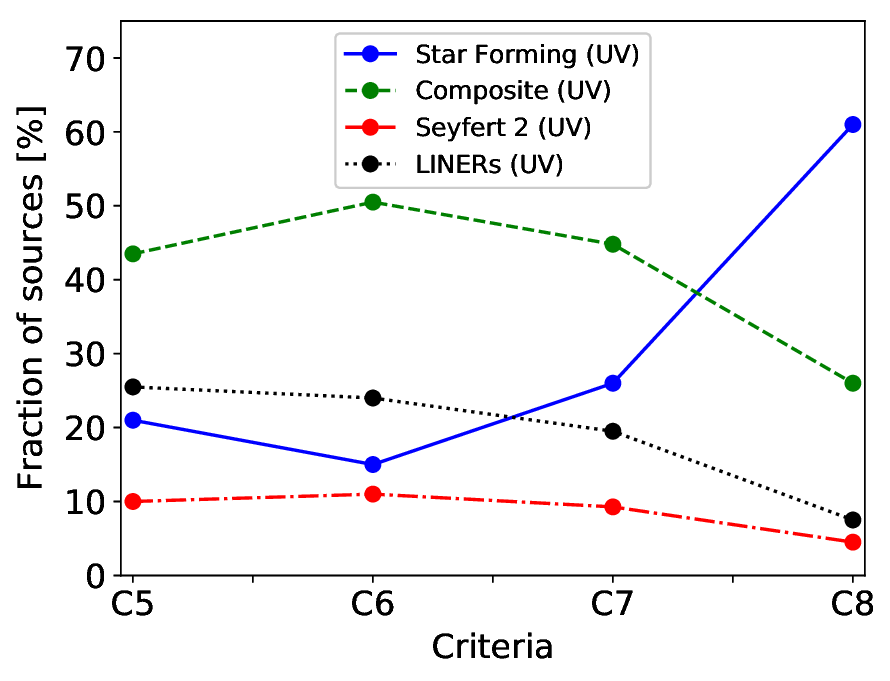}
\caption{The distribution of spectroscopic types for green valley galaxies selected using different optical (top) and UV (bottom) criteria. We present the star-forming galaxies (blue solid line), composite galaxies (green dotted line), Seyfert\,2 (red dash-dotted line), and LINERs (black dashed line).}
\label{fig-14}
\end{figure}

We analysed the selection of different spectroscopic types in the green valley when using different criteria in the optical and the UV. Using the BPT diagram ($\log$\,([\ion{O}{III}]$\lambda$5007/H$\beta$) vs. $\log$\,([\ion{N}{II}]/H$\alpha$)) defined in \citet{Baldwin}, we were able to classify the galaxies in our samples into star-forming, composite galaxies, Seyfert\,2, and LINERs. We selected sources with a signal-to-noise ratio S/N\,$>$\,3 in all four lines ([\ion{O}{III}]$\lambda$5007, H$\beta$, H$\alpha$, and [\ion{N}{II}]$\lambda$6584). We use the following references to previous works to separate the different classes: \cite{Kauffmann2003} to separate star-forming and composite galaxies: \cite{Kewley2001} to separate composite and AGN galaxies, and \cite{Scha} to separate Seyfert\,2 and LINERs.
\noindent In Table~\ref{table-4} we present the statistics for our spectroscopic classification, where the percentages in column 2 are calculated by comparing the number of sources obtained after considering the S/N\,$>$\,3 in all four lines with the number of original green valley samples. Fig.~\ref{fig-14} shows the comparison for both optically- (top panel) and UV-selected (bottom panel) samples.\\
\indent Most of the galaxies in the green valley are classified as star-forming or composite, in the case of most criteria. When comparing the optical and the UV criteria, the main difference is between C1 and C2, colour criteria without Gaussian fitting, with star-forming galaxies being the most selected in the C1 criterion and composite galaxies in the C2 criterion. For the colour criteria with Gaussian fit (C7 and C8), star-forming galaxies are dominant in both criteria without significant difference, as can be seen in Table~\ref{table-4}).

\begin{table}[ht!]
\scriptsize{
 \caption{Spectroscopic classification of the green valley selected samples into star-forming galaxies (SF), composites (Com), Seyfert\,2 (Sey), and LINERs.}
 \centering
 \begin{tabular}{l l l r r r} \hline\hline
 Criteria  & Parent& SF& Com& Sey&LINERs \\ %
    \hline
 C1(g\,-\,r) & 50483 (64\%)& 66\% &25.2\% & 5.3\% & 3.6\%     \\
 C2 (NUV\,-\,r) & 11459 (63\%) &21\%&  43.5\%& 10.0\%& 25.5\% \\
  C3 (SFR-M$_\ast$, optical) & 31157 (57\%)  & 30\%   &49.3\%  & 8.9\%& 11\%\\ 
  C4 (SFR-M$_\ast$, UV) & 10458 (59\%) &15\% &50.4\% &11\% & 24\%  \\
   C5 (sSFR, optical)  & 47830 (56\%) & 24\%    &45.5\% & 9.7\%  & 20\%        \\
   C6 (sSFR, UV) &22033 (71\%) & 26\% & 44.8\%  &9.3\%  &  19.5\%             \\
 C7 (g\,-\,r, Gauss) & 72578 (62\%)  & 67\%   &24\% & 5\% & 4\%    \\
 C8 (NUV\,-\,r, Gauss) & 46300 (84\%)  & 61\%   &26\%  & 4.5\%& 7.5\%         \\ 
  \hline\hline
  \end{tabular}%
  \label{table-4}
  }
 \end{table}

\subsection{Effect of extinction on green valley selection}\label{Extinction}

\begin{figure}[ht!]
\centering
\includegraphics[width=0.4\textwidth]{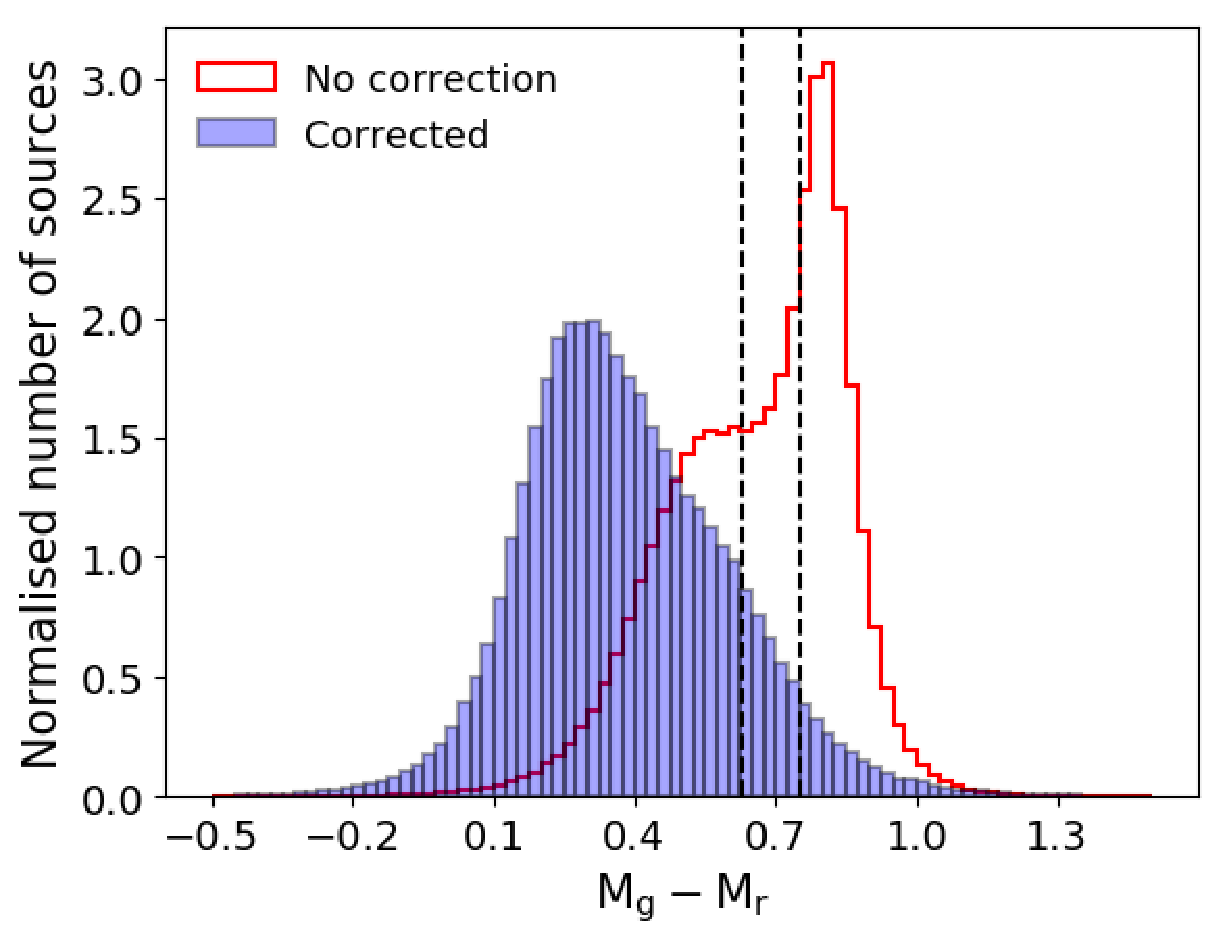}\\
 \includegraphics[width=0.4\textwidth]{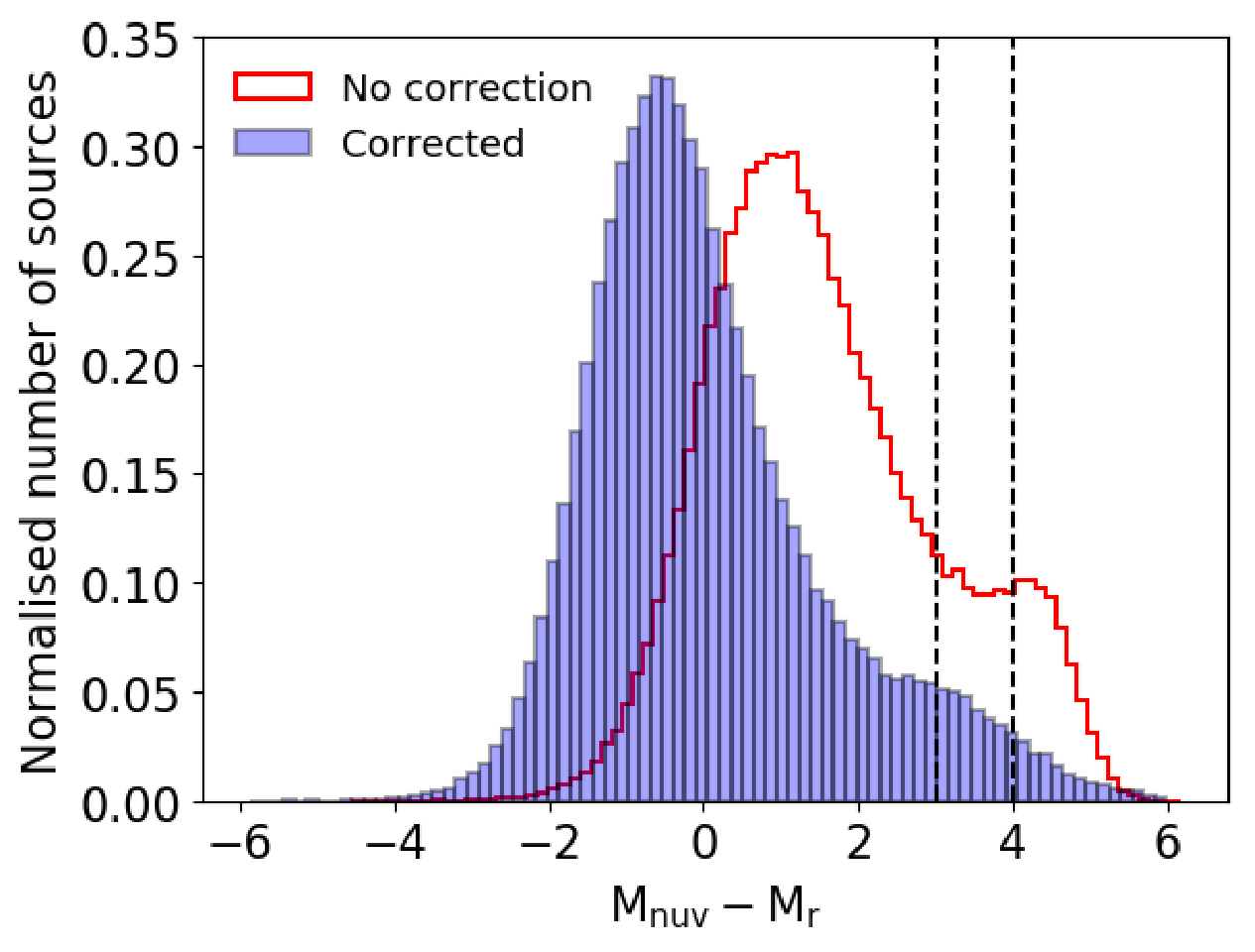}
\caption{M$_{\rm{g}}$-M$_{\rm{r}}$ (top panel) and M$_{\rm{NUV}}$-M$_{\rm{r}}$ (bottom panel) colour distributions before (red histograms) and after (blue semi-filled histograms) the extinction correction. The green valley region defined in Sections~\ref{colour(Opt)} and \ref{colour(UV)} is represented by two vertical black-dashed lines.}
\label{fig-222}
\end{figure}

The selection of the green valley using the colour criteria may be affected by dust extinction. Since the previously defined colour selection criteria (C1, C2, C7 and C8) did not consider any dust extinction, we study its possible effect in this section. One of the most reliable techniques to estimate dust extinction is through the H$_{\alpha}$/H$_{\beta}$ flux ratio (i.e. the Balmer decrement), which has been commonly used, in particular in the local Universe \citep{Dom}. We define the dust extinction from the Balmer decrements following the steps described in \cite{Mirjana2016}.\\
\indent Fig.~\ref{fig-222} shows the C1 (top) and C2 (bottom) criteria before (in red) and after (in blue) applying the dust extinction correction. As can be seen, the two distributions are
significantly different, both in the overall distribution and in the number of sources in the green valley. In particular, the number of sources is significantly different, and many sources are lost after applying the extinction correction because either H$\alpha$, H$\beta$, or both lines are not strong and/or not present in the SDSS galaxy spectra. Elliptical galaxies are more affected by the extinction correction than spirals, as they do not have strong emission lines (i.e., after correcting for extinction, we lose more elliptical galaxies than spiral galaxies). This will affect the bi-modal distribution of galaxies, as can be seen in Figure~\ref{fig-222} and thus the definition of the green valley for the extinction-corrected samples. \\
\indent Using the same green valley galaxy selection criteria as used in Sections~\ref{colour(Opt)} and \ref{colour(UV)}, we find that the extinction correction is affecting the green valley selection, as mentioned above. Using the M$_{\rm{g}}$-\,M$_{\rm{r}}$ criterion, we selected 28696 galaxies in the green valley (7.8\%), compared to 74663 (20\%) galaxies that were in the green valley before the extinction correction. There are 1649 (0.5\%) galaxies that were in the green valley before and remained in after correcting for extinction. Using the M$_{\rm{NUV}}$-\,M$_{\rm{r}}$ criterion, we now obtained 7546 galaxies in the green valley (4\%), compared to 18230 (10\%) galaxies that were in the green valley before correcting for extinction. We have 1146 (0.65\%) galaxies that were selected as green valley galaxies before and remained in the green valley after the extinction correction.

\section{Discussion}\label{discussion}

\begin{figure}
\centering
\includegraphics[width=0.4\textwidth]{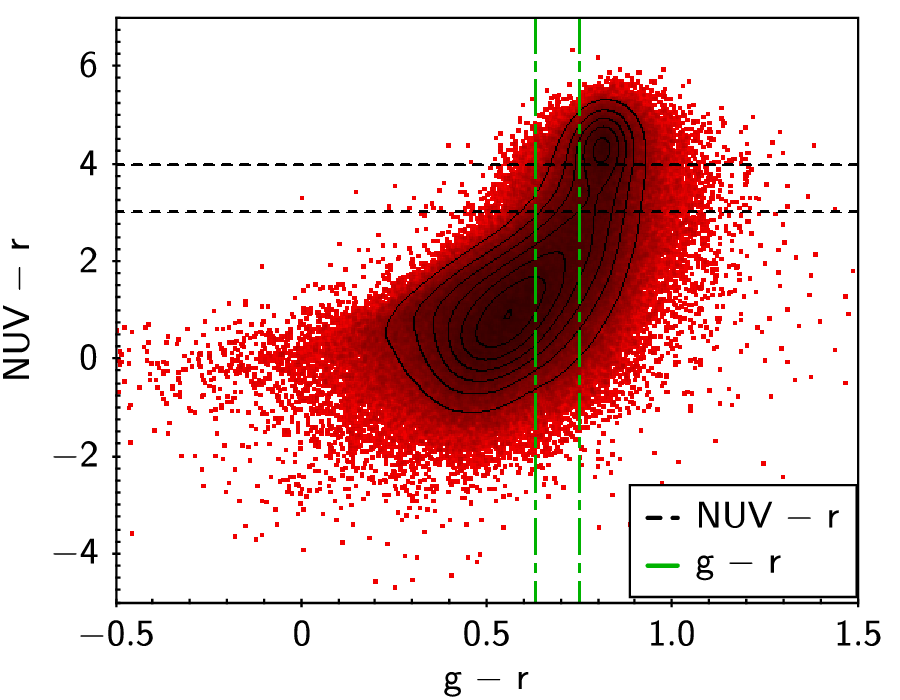}
\caption{The colour-colour criterion, with the possibility of green valley selection using either the g\,-\,r (green dot-dashed lines) or the NUV\,-\,r (black dashed lines) colours.}
\label{fig-1A}
\end{figure}

\begin{figure}
\centering
\includegraphics[width=0.4\textwidth]{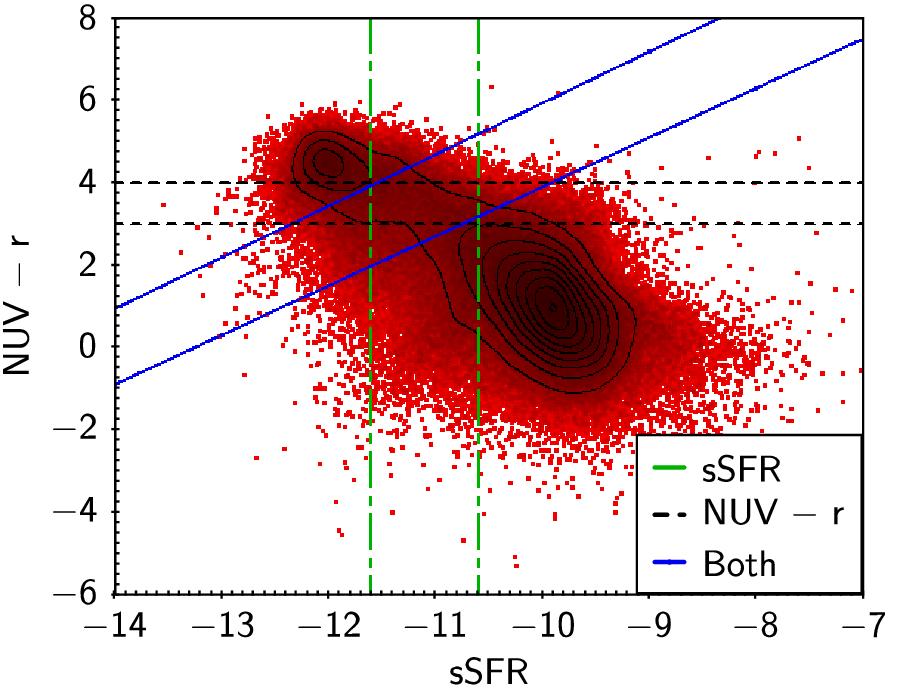}
\caption{NUV\,-\,r vs. sSFR relation, with the green valley selection using either the sSFR criterion (green dash-dotted lines), the NUV\,-\,r criterion (black dashed lines) or mixed criteria (blue solid lines).}
\label{fig-2A}
\end{figure}

Our results show that the selection of green valley using different criteria in optically- and UV-selected samples introduces differences in the number of selected sources and their properties in terms of stellar mass, SFRs, sSFRs, intrinsic brightness, morphological and spectroscopic types. In all cases, using the KS test, we find that the optical and UV samples have different properties. The largest difference in properties is associated with the colour criteria, in particular without applying Gaussian fittings, while the criteria based on the sSFR, and SFR vs. M$_\ast$ show smaller variations in the properties of the optical and UV samples. Therefore, we recommend the use of the sSFR or the SFR vs. M$_\ast$ criteria for the selection of the green valley. If these parameters are not available, and colours are used instead, we recommend selecting the green valley using Gaussian fittings of the bi-modal galaxy distribution, rather than a visually determined green valley, to avoid subjectivity.\\
\indent Overall, the fraction of green valley galaxies obtained in this work using the colour (without Gaussian fittings), sSFR, and SFR vs. M$_\ast$ criteria in the case of optical and UV samples is in line with previous results suggesting that green valley galaxies account for approximately 10\%\,-\,20\% of galaxies in all environments (e.g., \citealt{Schawinski, Jian2020, Das2021}). However, using the colour criteria with Gaussian fittings leads to a higher fraction of green valley galaxies of $\sim$\,30\% in the case of both optical and UV samples, as can be seen in Table~\ref{table-1}. The selection of the green valley by the Gaussian fitting can also change when considering different ranges of absolute magnitudes (or stellar masses), as shown by previous studies (e.g., \citealt{Wyder2007, Jin2014}).\\
\indent The stellar mass shows a slight difference between the optically- UV-selected green valley galaxies, with the UV sample having, on average, more massive galaxies compared to the optical samples using the same criteria, and a slightly higher
fraction of elliptical galaxies (see Table~\ref{table-2}). This is consistent with the findings of \cite{Sobral2018}, who suggested that the most massive O and B stars with lifetimes of a few Myr generate continuous UV radiation, and with the results of \cite{Sawicki}, who found a linear correlation between the stellar mass of galaxies and their UV luminosity. According to these studies, the most massive galaxies, being the main contributors to the aforementioned massive stars, may show a high UV luminosity. In addition, the selection of more massive galaxies in the UV than in the optical also leads to the selection of more galaxies that are
intrinsically brighter, as shown in Section~\ref{sec_mag_abs}.\\
\indent Our analysis was extended by classifying green valley samples morphologically and spectroscopically.
Morphological classification based on the Galaxy Zoo \citep{Lintott2008, Lintott} shows that spiral galaxies are more dominant than elliptical galaxies in all green valley selected samples, regardless of whether optically- or UV-selected data were used. This is in line with previous green valley studies, which reported a higher fraction of spiral galaxies than ellipticals \citep[e.g., see][and references therein]{Das2021}. It is also in line with previous suggestions that elliptical and spiral galaxies have significantly different evolutionary paths to and through the green valley, with elliptical galaxies transiting faster through the green valley than spirals, leading to a larger fraction of spiral galaxies \citep[e.g.,][]{Schawinski, Bremer2018, Bryukhareva2019, Quilley_Lapparent2022}. However, previous studies also show inconsistencies and broad range regarding the number of spiral and elliptical galaxies in the green valley, reporting spirals to be between 70\%\,-\,95\% and the elliptical galaxies to account for 5\%\,-\,30\%, depending on the study (e.g., \citealt{Salim2014, Bait2017, Lin2017, Das2021, Aguilar2025}). From our analysis, we can see that these inconsistencies in fractions among different studies can be explained by the use of either optical or UV-selected samples and different green valley criteria. As can be seen in Table~\ref{table-2}, depending on the criterion, we select 3\%-6\% (34\%-41\%) and 7\%-14\% (24\%-38\%) of elliptical (spiral) galaxies in the optical and UV, respectively, when considering the total sample, with many galaxies remaining unclassified in the Galaxy Zoo. However, if considering all classified galaxies, the above fractions become 7\%-15\% (85\%-93\%) and 16\%-37\% (63\%-84\%) of elliptical (spiral) galaxies in the optical and UV, respectively.\\
\indent We find different distributions in the SFRs (and thus also in the sSFRs) depending on the optical or UV samples and the criteria used, as discussed in Sections~\ref{sec_SFR} and \ref{sec_sSFR}. This can be explained by the selection of different fractions
of elliptical/spiral galaxies, depending on the criteria as shown above, but also by different fractions of spectroscopic types selected by each criterion. In particular, the selection of star-forming galaxies being dominant in C1, C7, and C8 (above 60\% in all cases), over the selection of composite galaxies being dominant in C3 and C5 in optical, and C2, C4, and C6 in the UV, with 40\%-50\%. The fraction of Seyfert 2 galaxies, $\sim$\,4\%-10\% is the lowest for almost all criteria, followed by LINERs from $\sim$\,4\%-25\% depending on the criterion. This could explain again the diversity of previous findings regarding the star formation properties of galaxies in the green valley \citep[e.g.,][]{Schiminovich2007, Damen, Pan2013, Schawinski, Chang, Coenda, Guo2021, Quilley_Lapparent2022}, suggesting different scenarios from slow quenching, extended quenching due to mergers and star formation, up to fast quenching due to starbursts, with and without significant impact of AGN feedback on star formation \citep[e.g.,][]{Zewdie2020, Kalinova2021}.\\\\
In addition to 1D and 2D colour-based and SFR-based methods discussed here, some studies have applied hybrid methods to select the green valley, suggesting that they could be more refined to identify green valley galaxies (e.g., \citealt{Salim2014}). This includes combining colour-based and/or SFR-based methods, such as colour-colour (e.g., NUV\,-\,r vs. g\,-\,r) or colour-sSFR (e.g., NUV\,-\,r vs. sSFR) diagrams. The underlying idea is that these hybrid methods can help to disentangle the effects of dust attenuation and young and old stellar populations, and thus reduce misclassification caused by dust reddening. However, these approaches require high-quality and well-calibrated photometric data to minimise uncertainties and ensure robust results (e.g., \citealt{Salim2014, Pandey2024}), and in addition, may decrease the number of sources in the sample and introduce more complexity in the selection of the green valley. As an example, we show in Figs.~\ref{fig-1A} and ~\ref{fig-2A}, the NUV\,-\,r vs. g\,-\,r and NUV\,-\,r vs. sSFR diagrams, respectively, for our sample. The lines drawn in the two figures show different possibilities of selecting the green valley according to the two galaxy overdensities, and thus the complexity of the two plots. Given all of the above, most previous studies have typically defined the green valley using a single criterion, either colour-based or SFR-based, rather than combining several diagnostics, which can lead to inconsistencies between studies (e.g., \citealt{Salim2014, Pandey2024}). Therefore, in this work, we only considered the C1-C8 criteria, which are standard and the most used in previous works. The analysis of other criteria could be the subject of future work.

\indent Overall, our results suggest that the green valley is not a single population but a diverse class of galaxies at different evolutionary stages, depending on the selection criterion used. This is in line with some of the previous studies (e.g., \citealt{Schawinski, Smethurst2015,  Bremer2018, Nogueira-Cavalcante2018, Das2021, Lin2022, Smith, Lin2024, Das2025}). To obtain a comprehensive view of the green valley galaxy evolution, both optical and UV selection methods should be combined, allowing researchers to study the full spectrum of quenching processes.

\section{Conclusions}\label{conclusion}

In this work, we have compared for the first time the eight most commonly used green valley selection criteria and analysed the properties of the selected galaxies and how they may affect the results reported in previous studies. To our knowledge, this is
the first study of its kind and the most detailed, which
should benefit future studies on green valley galaxies and
their selection. We used the SDSS optical and GALEX UV data at z\,$<$\,0.1 to analyse the properties of galaxies in different green valley samples selected based on optical and NUV colours (without and with Gaussian fittings), sSFR, and SFR vs. M$_{\ast}$ criteria and analysed the properties of selected galaxies from one criterion to another. The main findings are that green valley galaxies selected using different criteria in optical and UV may present different types of galaxies in terms of their M$_\ast$, SFR, sSFR, intrinsic brightness, morphology, and spectroscopic type. Therefore, this should be taken into account in all green valley studies that normally use only a single selection criterion when interpreting the results obtained. Our main results are as follows:
\begin{itemize}
    \item The stellar mass of the green valley galaxies is slightly higher, and the galaxies are intrinsically brighter when selected in the UV than in optical. They show very different SFRs and sSFRs depending on the criteria used. 
    \item In general, independently of optical and UV data, spirals are more dominant than elliptical galaxies in the green valley. However, the fraction of spiral and elliptical galaxies changes in optical (85\%-93\% and 7\%-15\%, respectively) and UV (63\%-84\% and 16\%-37\%, respectively), depending on the used criterion.
    \item Different criteria in the optical and the UV have tendencies of selecting either more star-forming or more composite galaxies, being followed by LINERs and finally Seyfert 2 galaxies. 
    \item The dust extinction can significantly affect the green valley selection based on colours in both optical and UV, in particular due to the loss of galaxies with fully absent or weak emission lines. 
    \item The criteria of green valley selection based on colours, g\,-\,r in optical and NUV\,-\,r in the UV, are the most sensitive to the galaxy properties. We found other sample selection criteria based on sSFR and the SFR vs. M$_{\ast}$ to be more stable, and more recommendable to use. If the colour criteria are to be used, we recommend selecting the green valley through the Gaussian fitting of the bi-modal distribution of galaxies, rather than visually selecting the green valley to avoid subjectivity.
\end{itemize}
\indent Finally, our results suggest that the green valley is not a single population but a diverse class of galaxies at different evolutionary stages, depending on the sample and the selection criterion used. To obtain a complete view of galaxy evolution through the green valley, both optical and UV selection methods should be combined, allowing the study of the full spectrum of quenching processes.

\section*{Data availability}
\small{This work is predominantly based on the public data from the Sloan Digital Sky Survey (SDSS) available at \url{https://wwwmpa.mpa- garching.mpg.de/SDSS/DR7/}, GALEX data from the Galaxy Evolution Explorer All Sky Imaging Surveys (GALEX-AIS) available at \url{https://archive.stsci.edu/pub/hlsp/bianchi_gr5xdr7/}. And data from Galaxy Zoo catalogue available at \url{https://data.galaxyzoo.org/}.} 
\section*{Acknowledgements}
\small{We thank the anonymous referee for all his/her comments, which have improved our manuscript. Financial support from the Swedish International Development Cooperation Agency (SIDA) through the International Science Programme (ISP)-Uppsala University to the University of Rwanda through the Rwanda Astrophysics, Space and Climate Science Research Group (RASCSRG), and through the East African Astrophysics Research Network (EAARN) is gratefully acknowledged. MP acknowledges financial support from the Spanish MCIU through the project PID2022-140871NB-C21 by “ERDF A way of making Europe” and the project PID2024-162972NB-I00, the Severo Ochoa grant CEX2021-515001131-S funded by MCIN/AEI/10.13039/501100011033, and the support from the Space Science and Geospatial Institute (SSGI) under the Ethiopian Ministry of Innovation and Technology (MInT). AM acknowledges support from the National Research Foundation of South Africa. This research made use of several Python libraries, including NumPy \citep{Harris2020}, SciPy \citep{Virtanen2020}, Matplotlib \citep{Hunter2007}, and Astropy (\citealt{Astropy2013, Astropy2018}), in addition to the Jupyter Notebook.}\\\\

\bibliographystyle{aa}
\bibliography{Bibliography} 




\label{lastpage}
\end{document}